\documentclass{IEEEtaes}
\usepackage{color,array,amsthm}
\usepackage{graphicx}
\usepackage{amsmath,amssymb,amsfonts}
\usepackage{algorithmic}
\usepackage{array}
\usepackage[caption=false,font=normalsize,labelfont=sf,textfont=sf]{subfig}
\usepackage{textcomp}
\usepackage{stfloats}
\usepackage{url}
\usepackage{verbatim}
\usepackage{graphicx}
\usepackage{algorithm}
\usepackage{color}
\usepackage{xcolor}
\usepackage{booktabs}
\usepackage{multirow}
\usepackage{hyperref}
\usepackage{pifont} 
\usepackage{colortbl}
\usepackage{cite}
\usepackage{makecell}
\usepackage{orcidlink}

\newtheorem{proposition}{PROPOSITION}
\newtheorem{corollary}{COROLLARY}
\newtheorem{lemma}{LEMMA}
\newtheorem{remark}{REMARK}

\jvol{XX}
\jnum{XX}
\jmonth{XXXXX}
\paper{1234567}
\pubyear{2020}
\doiinfo{TAES.2020.Doi Number}

\begin{document}	
\title{Label-Free Performance Estimation of Deep Learning Based Radar Detectors in Time-Varying Environments}

\author{Chuanfei Zang\orcidlink{0009-0005-8851-5926}}
\member{Graduate Student Member, IEEE}

\author{Yiru Lin\orcidlink{0009-0004-6850-5344}}
\member{Graduate Student Member, IEEE}

\author{Yumiao Wang\orcidlink{0000-0002-1474-0486}}
\member{Member, IEEE}

\author{Xingyu Chen\orcidlink{0009-0001-4295-9928}}
\member{Graduate Student Member, IEEE}

\author{Xiaobo Yang\orcidlink{0009-0000-4888-4050}}
\member{Member, IEEE}
\affil{University of Electronic Science and Technology of China, Chengdu, China}

\author{Guolong Cui\orcidlink{0000-0001-5707-6311}}
\member{Senior Member, IEEE}
\affil{University of Electronic Science and Technology of China, Chengdu, China\\Tianfu Jiangxi Laboratory, Chengdu, China}

\receiveddate{This work was supported by the National Natural Science Foundation of China under Grants U24B20188 and 62271126, and by the Sichuan Province Innovative Talent Funding Project for Postdoctoral Fellows.}
\authoraddress{Authors’ addresses: Chuanfei Zang, Yiru Lin, Yumiao Wang, Xingyu Chen, and Xiaobo Yang are with the School of Information and Communication Engineering, University of Electronic Science and Technology of China, Chengdu 611731, China (e-mail: zangcf2017@163.com; linyiru2021@163.com; wangyumiao@uestc.edu.cn; cxytoby@163.com; yangxb@uestc.edu.cn). Guolong Cui is with the School of Information and Communication Engineering, University of Electronic Science and Technology of China, Chengdu 611731, China, and also Tianfu Jiangxi Laboratory, Chengdu, Sichuan 641419, China (e-mail: cuiguolong@uestc.edu.cn).(\textit{Corresponding author: Yumiao Wang}.)}

\markboth{ZANG ET AL.}{LABEL-FREE PERFORMANCE ESTIMATION}
\maketitle
	
\begin{abstract}
	Deep-learning-based radar target detectors can achieve strong performance in complex sensing environments. However, in time-varying environments, changes in clutter characteristics may cause the distribution of echoes to deviate from that of the training data, thereby degrading detector reliability. Timely identification of such degradation requires an accurate assessment of the deployed detector's performance; however, ground-truth labels for radar echoes are typically unavailable during deployment. This paper investigates label-free performance estimation for deployed deep learning-based radar target detectors. We first show that confidence-based estimators tend to overestimate detection performance under distribution shift because they fail to account for the discrepancy between the detector-induced pseudo-label distribution and the true label distribution. To address this limitation, we propose probabilistic optimal transport (POT), which assesses deployment performance by computing the Wasserstein distance between the detector's output probability distribution on unlabeled deployment echoes and the label distribution in the training data. We further develop thresholded probabilistic optimal transport (TPOT), which suppresses the influence of anomalous transport costs on the Wasserstein-distance-based estimate, thereby improving estimation robustness. Experiments on a publicly measured radar dataset demonstrate that POT and TPOT consistently outperform representative label-free baselines across deployment scenarios with varying degrees of data distribution shift.
\end{abstract}
	
\section{Introduction}
Radar target detection aims to reliably determine the presence of target echoes against noise and clutter and is a fundamental task in radar signal processing~\cite{10535463,10675372,11543476}. In complex sensing environments, heterogeneous strong clutter and variations in background scattering can substantially weaken the discriminative characteristics of target echoes, posing severe challenges to detector reliability~\cite{10371148,10848402,10929070,10879091}.

Conventional radar detectors are generally designed based on prespecified statistical clutter models, and their performance therefore depends strongly on the agreement between the assumed model and the actual clutter distribution~\cite{9807335,10742946,11193878}. In practice, clutter often exhibit complex heterogeneous, nonstationary, and non-Gaussian statistics that cannot be accurately characterized by prespecified models. Consequently, the performance of conventional detection methods is markedly limited in complex environments~\cite{11193878,11267251,9145700}.

In recent years, deep learning-based radar target detectors have outperformed conventional methods in complex scenarios because of their strong feature-representation and nonlinear-modeling capabilities~\cite{10415070,10371351,11511344,10530230}. These methods, however, generally assume that the training and deployment data are identically or approximately identically distributed~\cite{11425796,11125835}. During deployment, changes in clutter conditions and operating environments may induce shifts in the radar echo distribution, thereby creating a mismatch between the deployment and training distributions and degrading detection performance~\cite{11328771,9622244}.

Existing studies have sought to restore detector performance in deployment scenarios through adaptive learning or self-evolving training~\cite{10587212,10975280}. However, updating a deployed detector typically requires additional computational resources, continuous data processing, and reliable supervision, while unnecessary updates may increase maintenance costs and cause error accumulation due to low-quality samples or incorrect pseudo-labels~\cite{11359362,11124398,10848058,11231088}. It is therefore preferable to estimate the operating performance of the deployed detector and trigger detector adaptation or retraining only when substantial performance degradation is identified~\cite{8496795,11103500,7450691}. Performance estimation thus provides a critical link between deployment-time performance monitoring and detector adaptation.

Ground-truth labels for radar echo are rarely available in a timely manner during deployment. Noncooperative targets are difficult to verify in real time because of their uncertain occurrence, trajectories, and locations, while radar echo annotation typically requires manual verification, multisource corroboration, or post hoc analysis, resulting in substantial cost and latency~\cite{10066514,9655472}. Therefore, this paper focuses on reliably estimating the performance and degradation of a deployed deep learning-based radar target detector using only unlabeled radar echoes from the current sensing scenario.

Existing label-free methods for estimating the performance of deployed detectors can be broadly categorized into post-deployment training, pre-deployment training, and output-driven methods~\cite{9655472}. Post-deployment training methods train a new auxiliary detector using unlabeled deployment data and the corresponding pseudo-labels, and estimate the performance degradation of the deployed detector based on the parameter changes between the auxiliary and deployed detectors~\cite{yu2022predicting}. These methods require additional training for each batch of deployment data, incur substantial computational cost, and are sensitive to label quality, sample size, and the optimization process~\cite{10858421}.

Pre-deployment training methods train an additional detector in advance, typically using the same architecture and source training data as the deployed detector but an independent optimization run, and estimate deployment performance from the prediction disagreement between the two detectors on unlabeled deployment data~\cite{jiang2022assessing}. However, this approach requires additional training and model storage, and its reliability depends on sufficient predictive diversity between the two detectors. Because the detectors share the same training data, architecture, and learning procedure, they may exhibit correlated errors under distribution shift. Errors on which both detectors produce the same incorrect prediction are not captured by their disagreement, potentially leading to an underestimation of the deployment error rate and, consequently, an overestimation of detector performance\cite{pagliardini2023agree,kirsch2022note}.

By contrast, output-driven methods estimate the performance of a deployed detector solely from its predictions on unlabeled deployment data, without requiring additional model training, and are therefore well suited to online detector performance monitoring. A representative method is average confidence (AC), which estimates the detection accuracy of the deployed detector using the mean confidence of its predictions on the deployment data~\cite{ovadia2019can}. Although AC is simple and computationally efficient, its reliability requires the detector's predictive confidence to remain well calibrated on the deployment distribution, such that the mean confidence provides an accurate approximation to the actual detection accuracy. This requirement may be violated when the distribution of radar echoes in the deployment environment deviates substantially from the training distribution, because calibration established on the training data does not necessarily transfer to the shifted deployment data. Consequently, AC may produce large errors in the estimated detector performance~\cite{ovadia2019can,lipton2018detecting}.

Motivated by this limitation, this paper investigates the reliability of output-driven performance estimation under distribution shift. We first analyze AC using optimal transport (OT) theory~\cite{ovadia2019can}. The analysis shows that AC-type methods tend to overestimate the actual performance of deployed detectors under distribution shift and that the degree of overestimation is closely related to shifts in the detector-induced pseudo-label distribution~\cite{chen2021mandoline}. Pseudo-label shift refers to the discrepancy between the predicted-label and ground-truth-label distributions and reflects the detector's overall decision bias in the current sensing scenario. AC cannot adequately capture the estimation bias associated with pseudo-label shift because it only averages the predictive confidence produced by the detector across radar echo samples, without characterizing the redistribution of predicted probability mass among the clutter and target classes.

These observations indicate that reliable label-free performance estimation under distribution shift should account for the classwise redistribution of predictive probability mass and compare the resulting output distribution with a reference distribution that is independent of the detector-induced pseudo-labels. OT provides a natural framework for this purpose, as its transport cost quantifies the probability mass required to reconcile two distributions over a common space. Accordingly, we propose a probabilistic optimal transport (POT) performance estimation method. POT constructs an empirical probability distribution from the class probabilities output by the detector on current unlabeled echoes and computes its Wasserstein distance to the empirical training-set label distribution. This distance measures the deviation of the detector's output distribution from the training-stage label distribution and is used to estimate performance degradation in the current scenario. Unlike methods based solely on samplewise confidence statistics, POT preserves the classwise structure of the predictive probability mass and performs global distribution matching against the reference label distribution, thereby enabling it to capture systematic decision shifts that may be obscured by high-confidence predictions.

We further analyze the performance estimation properties of POT and derive a corresponding lower bound. The theoretical results show that POT provides accurate performance estimates even under substantial pseudo-label shift. To further reduce the influence of anomalous outputs on the OT cost, we develop a robust POT variant termed thresholded probabilistic optimal transport (TPOT). TPOT introduces a learnable thresholding mechanism that adaptively suppresses the contribution of anomalous samples to the transport cost, thereby improving the robustness of POT in complex deployment scenarios. The main contributions of this paper are summarized as follows:
\begin{enumerate}
	\item Using OT theory, we analyze why AC-type methods tend to overestimate detector performance under distribution shift and establish the relationship between the resulting estimation bias and pseudo-label shift.
	\item Motivated by this analysis, we propose POT, which replaces the detector-induced pseudo-label distribution with the empirical training-set label distribution as the reference and measures the resulting Wasserstein distance, thereby enabling more reliable performance estimation.
	\item We theoretically analyze the performance estimation properties of POT and derive a corresponding lower bound, demonstrating that POT provides accurate performance estimates even under substantial pseudo-label distribution shifts in the deployment data.
	\item We develop TPOT, a robust variant of POT that suppresses the influence of anomalous outputs on the OT cost through a learnable thresholding mechanism, further improving estimation robustness under complex distribution shifts.
\end{enumerate}

The remainder of this paper is organized as follows. Section~\ref{sec:problem_formulation_and_related_work} formulates the problem and reviews related work. Section~\ref{sec:Preliminaries} introduces OT theory and presents the relevant theoretical analysis. Section~\ref{sec:method} describes the proposed label-free performance estimation methods. Section~\ref{sec:experiments} reports and analyzes the experimental results. Finally, Section~\ref{sec:conclusion} concludes the paper.
	
\section{Problem Formulation and Related Work}\label{sec:problem_formulation_and_related_work}
This section first establishes the radar signal model in Section~\ref{subsec:Signal_Model}, formulates the label-free performance estimation problem in Section~\ref{subsec:Problem_Formulation}, and then reviews output-driven performance estimation methods and analyzes their principles and limitations in Section~\ref{subsec:Output-Driven_Estimation}.

\subsection{Signal Model}\label{subsec:Signal_Model}
Within a coherent processing interval (CPI), the radar transmits $P$ coherent pulses and receives the corresponding echoes through the in-phase/quadrature (I/Q) channels. After quadrature demodulation, matched filtering, and range processing, a discrete echo matrix $\mathbf{X} \in \mathbb{C}^{P \times R}$ is obtained, where $R$ is the number of range cells and $\mathbf{x}_r \in \mathbb{C}^{P \times 1}$ is the slow-time echo vector associated with the $r$-th range cell. For an arbitrary cell under test (CUT), the received echo can be described by the following binary hypothesis testing model:
\begin{equation}
	\begin{cases}
		H_0:\mathbf{x}_r=\mathbf{c}_r+\mathbf{n}_r,\\
		H_1:\mathbf{x}_r=\mathbf{s}_r+\mathbf{c}_r+\mathbf{n}_r,
	\end{cases}
	\label{eq:target}
\end{equation}
where $\mathbf{s}_r$, $\mathbf{c}_r$, and $\mathbf{n}_r$ denote the target signal component, clutter signal, and additive noise, respectively. Hypotheses $H_1$ and $H_0$ correspond to target presence and target absence, respectively; the associated CUTs are referred to as target and clutter cells.

In addition to determining target presence, a deep learning-based detector may further identify the type of the target present in the CUT. To describe target detection and target-type identification within a unified framework, different target types and the clutter background are treated as mutually exclusive classes, and the detection task is formulated as a $K$-class classification problem. One class represents the clutter background, whereas the remaining $K-1$ classes correspond to different target types. When only target presence is considered and the specific target type is irrelevant, the model reduces to binary classification with $K=2$.

Let $f_{\boldsymbol{\theta}}$ denote a deep learning-based detector parameterized by $\boldsymbol{\theta}$ that maps an input echo $\mathbf{x}_r$ to the $\left( K-1\right) $-dimensional probability simplex $\Delta^{K-1}$:
\begin{equation}
	f_{\boldsymbol{\theta}}\!:\!\mathcal{X}\!\to\!\Delta^{K\!-\!1}\!,
	\Delta^{K\!-\!1}\!\!=\!\left\{\!\mathbf{p}\!\in\!\mathbb{R}^{K}\middle| \mathbf{p}_k\!\ge 0,\! \sum_{k=1}^{K}\!\mathbf{p}_k \!=\!1\!\right\}.
	\label{eq:model_output}
\end{equation}
The $k$-th component $[f_{\boldsymbol{\theta}}(\mathbf{x}_r)]_k$ of the detector output $f_{\boldsymbol{\theta}}(\mathbf{x}_r)$ is interpreted as the predicted probability that the input sample $\mathbf{x}_r$ belongs to class $k$. The detection result is then obtained using the maximum a posteriori criterion:
\begin{equation}
	\mathrm{dec}\!\left(f_{\boldsymbol{\theta}}(\mathbf{x}_r)\right)=\arg\max_{{k\in\{1,\ldots,K\}}}[f_{\boldsymbol{\theta}}(\mathbf{x}_r)]_k.
\end{equation}
Because all CUTs are processed separately using the same procedure, the range-cell index $r$ is omitted hereafter for notational simplicity, and $\mathbf{x}$ denotes the echo vector of the CUT under consideration.

\subsection{Problem Formulation}\label{subsec:Problem_Formulation}
Because the detector $f_{\boldsymbol{\theta}}$ is implemented by a deep neural network, its parameters $\boldsymbol{\theta}$ must be learned from labeled training data. Let the training and validation sets be
\begin{equation}
	\begin{aligned}
		&\mathcal{D}_{\mathrm{train}} = \{(\mathbf{x}^{(i)}_{\mathrm{train}}, \mathbf{y}^{(i)}_{\mathrm{train}})\}_{i=1}^{N_{\mathrm{train}}} \overset{\mathrm{i.i.d.}}{\sim} P_\mathrm{T}\left( \mathbf{x}, \mathbf{y}\right), \\
		&\quad\mathcal{D}_{\mathrm{val}} = \{(\mathbf{x}^{(i)}_{\mathrm{val}}, \mathbf{y}^{(i)}_{\mathrm{val}})\}_{i=1}^{N_{\mathrm{val}}} \overset{\mathrm{i.i.d.}}{\sim} P_\mathrm{T}\left( \mathbf{x}, \mathbf{y}\right),
	\end{aligned}
\end{equation}
where the samples are independently and identically drawn from the training-stage joint distribution $P_{\mathrm{T}}(\mathbf{x},\mathbf{y})$. The training set is used to learn the detector parameters, and the validation set is used for model selection and hyperparameter determination. $N_{\mathrm{train}}$ and $N_{\mathrm{val}}$ denote the numbers of training and validation samples, respectively, and $i$ is the sample index.

Class labels are represented by one-hot vectors. For sample $\mathbf{x}^{(i)}$ with ground-truth class $k$, its label $\mathbf{y}^{(i)}\in\{0,1\}^{K}$ satisfies
\begin{equation}
	{\mathbf{y}}^{(i)} =
	\begin{cases}
		{\mathbf{y}}^{(i)}_j = 1, \quad j = k ;\\
		{\mathbf{y}}^{(i)}_j = 0, \quad j \in \{1,...,K\} \setminus \{k\}.
	\end{cases}
\end{equation}

After model training, the deployed detector $f_{\boldsymbol{\theta}^{\ast}}$ is obtained. Let $P_\mathrm{D}\left( \mathbf{x}, \mathbf{y}\right)$ denote the data distribution in the deployment environment. Changes in clutter conditions and operating environments may cause the deployment distribution $P_\mathrm{D}\left( \mathbf{x}, \mathbf{y}\right)$ to deviate from the training distribution $P_\mathrm{T}\left( \mathbf{x}, \mathbf{y}\right)$. This distribution shift can reduce the applicability of the feature representations and decision boundaries learned by $f_{\boldsymbol{\theta}^{\ast}}$ during training to the current scenario, thereby degrading performance.

If the ground-truth labels of the deployment data were available, the detection error rate of the detector in the current deployment environment could be calculated as
\begin{equation}
	\mathcal{E}_{\mathrm{D}}(f_{\boldsymbol{\theta}^{\ast}}) = \mathbb{E}_{(\mathbf{x},\mathbf{y})\sim P_{\mathrm{D}}}
	\left[ \mathbb{I}\left( \mathrm{dec}(f_{\boldsymbol{\theta}^{\ast}}(\mathbf{x})) \neq \mathrm{dec}(\mathbf{y}) \right) \right],
	\label{eq:Error_detection_rate}
\end{equation}
where $\mathbb{I}(\cdot)$ is the indicator function. The corresponding detection accuracy is $1-\mathcal{E}_{\mathrm{D}}(f_{\boldsymbol{\theta}^{\ast}})$. This paper uses the deployment error rate as the primary metric of the detector's current operating performance. For notational simplicity, $\mathcal{E}_{\mathrm{D}}(f_{\boldsymbol{\theta}^{\ast}})$ is abbreviated as $\mathcal{E}_{\mathrm{D}}$ hereafter.

In practical deployment, however, the ground-truth label $\mathbf{y}_{\mathrm{dep}}$ associated with each echo data sample is generally unobservable, and the performance estimator has access only to unlabeled radar echo data from the deployment environment:
\begin{equation}
	\mathcal{D}_{\mathrm{dep}} = \{(\mathbf{x}^{(i)}_{\mathrm{dep}})\}_{i=1}^{N_{\mathrm{dep}}} \overset{\mathrm{i.i.d.}}{\sim} P_\mathrm{D}\left( \mathbf{x}\right).
\end{equation}	
Consequently, the deployment error rate $\mathcal{E}_{\mathrm{D}}$ cannot be directly evaluated from the unlabeled deployment data. Performance estimation during deployment can therefore be viewed as label-free error-rate estimation: given only the trained detector $f_{\boldsymbol{\theta}^{\ast}}$, the training-stage echo data, and the unlabeled deployment data $\mathcal{D}_{\mathrm{dep}}$, construct an error-rate estimator $g$ that produces the estimate
\begin{equation}
	{\mathcal{E}} = g\left(f_{\boldsymbol{\theta}^{\ast}}, \mathcal{D}_{\mathrm{dep}}, 	\mathcal{D}_{\mathrm{train}}, \mathcal{D}_{\mathrm{val}} \right),
\end{equation}
such that the estimate ${\mathcal{E}}$ approximates the true deployment error rate $\mathcal{E}_{\mathrm{D}}$ of the detector under the current deployment distribution as closely as possible.

\subsection{Related Work}\label{subsec:Output-Driven_Estimation}
Among existing methods for estimating detector performance from prediction outputs on unlabeled deployment data, AC is one of the most fundamental and widely used approaches~\cite{ovadia2019can}. It approximates classification accuracy using the mean predictive confidence of the detector on deployment data. For a deployment dataset $\mathcal{D}_{\mathrm{dep}}$, the AC error-rate estimator is
\begin{equation}
	{\mathcal{E}}_{\mathrm{AC}} = \mathbb{E}_{\mathbf{x}\sim \mathcal{D}_{\mathrm{dep}}} \left[ 1- \max\left( f_{\boldsymbol{\theta}^{\ast}}(\mathbf{x}) \right)   \right].
	\label{eq:AC_Estimate}
\end{equation}
where $\max\left( f_{\boldsymbol{\theta}^{\ast}}(\mathbf{x}) \right)$ is the largest component of the detector's output probability vector and is termed the predictive confidence.

Although AC is simple and requires no additional hyperparameters, its reliability depends strongly on the probabilistic calibration of the detector, i.e., whether its predictive confidence agrees with the actual probability of a correct prediction~\cite{donmez2010unsupervised}. For example, for a well-calibrated detector, a set of echo data samples with an average predictive confidence of $0.9$ should have an actual classification accuracy close to $90\%$. However, when the distribution of deployment radar echoes deviates from the training distribution, detectors tend to produce overconfident predictions and may assign high confidence to incorrect classes~\cite{10356834}. Consequently, the average predictive confidence may exceed the actual accuracy, causing AC to systematically underestimate the true deployment error rate and overestimate deployment performance. Moreover, averaging confidence scores can be affected by extreme values, resulting in unstable estimates.

To reduce sensitivity to numerical bias in directly averaged confidence scores, average thresholded confidence (ATC)~\cite{garg2022leveraging} converts numerical confidence statistics into threshold decisions. It first computes the detector error rate on the validation set:
\begin{equation}
	{\mathcal{E}}_{\mathrm{val}} \!=\! \frac{1}{N_{\mathrm{val}}}\! \sum_{i=1}^{N_{\mathrm{val}}}
	\mathbb{I} \left( \mathrm{dec} \left( f_{\boldsymbol{\theta}^{\ast}}\! \left(\mathbf{x}_{\mathrm{val}}^{(i)}\right) \right) \!\neq\! \mathrm{dec} \left(\mathbf{y}_{\mathrm{val}}^{(i)}\right) \right).
\end{equation}
A threshold $t^{\ast}$ is then selected on the validation set such that the proportion of samples whose reliability scores fall below the threshold is as close as possible to the validation error rate:
\begin{equation}
	t^{\ast} \!=\! \arg\min_{t\in\mathbb{R}} \left| \frac{1}{N_{\mathrm{val}}}\! \sum_{i=1}^{N_{\mathrm{val}}} \!\mathbb{I}\!
	\left( s\!\left( f_{\boldsymbol{\theta}^{\ast}}\! \left(\mathbf{x}_{\mathrm{val}}^{(i)}\right)\! \right) \!<\!t \right) \!-\! {\mathcal{E}}_{\mathrm{val}} \right|\!,
\end{equation}
where $s(\cdot)$ denotes a score function for quantifying prediction reliability, with a larger score indicating higher reliability; common choices include maximum confidence and negative entropy. Finally, the deployment error rate is estimated as the proportion of samples in the deployment data $\mathcal{D}_{\mathrm{dep}}$ whose reliability scores are below $t^{\ast}$:
\begin{equation}
	{\mathcal{E}}_{\mathrm{ATC}} =
	\mathbb{E}_{\mathbf{x}\sim \mathcal{D}_{\mathrm{dep}}}
	\left[ \mathbb{I}\left(s(f_{\boldsymbol{\theta}^{\ast}}(\mathbf{x}))<t^{\ast}\right) \right].
	\label{eq:ATC_Estimate}
\end{equation}
Compared with AC, which directly averages predictive confidence, ATC counts only the proportion of samples below a threshold. It can therefore reduce the influence of extreme confidence values and sensitivity to calibration errors in the absolute confidence scale. However, ATC still relies on the relationship between reliability scores and prediction errors remaining stable from validation to deployment. When distribution shift changes the relationship between reliability scores and prediction errors, ATC may still exhibit substantial estimation bias~\cite{kiyasseh2024framework}.

Furthermore, without additional constraints on the unknown deployment echo data distribution, unlabeled data and detector outputs alone generally do not uniquely determine the model's true performance in the deployment scenario~\cite{lu2023characterizing}. AC and ATC therefore require the training and deployment distributions to share the same conditional label distribution, i.e., the covariate-shift assumption
\begin{equation}
	{P}_{\rm D}(\mathbf{y}|\mathbf{x})={P}_{\rm T}(\mathbf{y}| \mathbf{x}).
\end{equation}
This assumption is difficult to satisfy strictly in practical radar deployment. Changes in the clutter environment and sensing conditions may alter not only the marginal distribution $P(\mathbf{x})$ but also the class-conditional distribution $P(\mathbf{x}|\mathbf{y})$ and posterior distribution $P(\mathbf{y}|\mathbf{x})$. Consequently, the relationship between predictive confidence and the error rate, as established on the source echo data, may not hold during deployment, thereby reducing the accuracy of the estimates produced by AC and ATC in complex radar scenarios~\cite{10452835}.

\section{Optimal Transport and Average Confidence}\label{sec:Preliminaries}
This section first reviews the fundamentals of OT and then analyzes the geometric interpretation of the AC error-rate estimator from an OT perspective. We show that the AC estimate of the deployed detector's error rate is equal to the Wasserstein distance between the detector's output probability distribution and the pseudo-label distribution induced by its predictions. We further characterize the relationship between pseudo-label shift and the underestimation of the detector's error rate by AC. For notational simplicity, the deployed detector $f_{\boldsymbol{\theta}^{\ast}}$ is denoted by $f$ hereafter. Proofs of all propositions and corollaries are provided in the appendices.

\subsection{Optimal Transport Preliminaries}
OT measures the discrepancy between two distributions by computing the minimum cost required to move probability mass from one distribution to the other. Let $\mu$ and $\nu$ be probability distributions defined on spaces $X$ and $Y$, respectively, and let $c(x,y)$ denote the cost of transporting a unit mass from source location $x\in X$ to target location $y\in Y$. Under the Kantorovich OT formulation~\cite{10476763}, a transport plan $\pi$ is a joint probability distribution on the product space ${X}\times{Y}$ satisfying the marginal constraints
\begin{equation}
	\pi(\cdot,Y)=\mu(\cdot), \quad \pi(X,\cdot)=\nu(\cdot).
\end{equation}
The transport plan $\pi$ describes the allocation of mass between the source and target distributions. In the discrete case, $\pi(x,y)$ denotes the probability mass transported from source location $x$ to target location $y$; when $\mu(x)>0$, $\pi(x,y)/\mu(x)$ is the fraction of the mass at $x$ assigned to $y$.

Let $\Pi(\mu,\nu)$ denote the set of all transport plans satisfying these marginal constraints. Given a transport cost function $c(x, y)$, the corresponding OT cost is defined as
\begin{equation}
	W_c(\mu, \nu) = \inf_{\pi \in \Pi(\mu, \nu)} \int_{X \times Y} c(x, y) d\pi(x, y) .
\end{equation}
When $c(x, y)$ is induced by an underlying metric, $W_c$ is the corresponding Wasserstein distance. Unlike total variation distance (TVD), which measures discrepancies in probability mass without accounting for the geometry of the underlying space, Wasserstein distance incorporates the cost of transporting probability mass between locations. It therefore captures both the discrepancy in probability-mass allocation between the two distributions and the geometry of transporting mass from one distribution to the other~\cite{gibbs2002choosing}.

For discrete probability measures, such as equally weighted empirical distributions, the Wasserstein distance reduces to the following linear programming problem:
\begin{equation}
	W_c(\mu,\nu) \!=\!\!\!\! \min_{\mathbf{P}\in\Pi(\mathbf{\mu},\mathbf{\nu})}
	\langle\mathbf{P},\mathbf{C}\rangle \!=\!\!\!\! \min_{\mathbf{P}\in\Pi(\mathbf{\mu},\mathbf{\nu})}
	\sum_{i=1}^{m}\!\sum_{j=1}^{n}\mathbf{P}_{ij}\mathbf{C}_{ij},
	\label{eq:discrete_ot}
\end{equation}
where $\mathbf{C}\in\mathbb{R}^{m\times n}$ is the cost matrix with $C_{ij}=c(\mu_i,\nu_j)$, and $\mathbf{P}\in\mathbb{R}_{+}^{m\times n}$ is the transport plan, where $P_{ij}$ denotes the probability mass transported from $\mu_i$ to $\nu_j$. When $\mu$ and $\nu$ contain the same number of equally weighted samples, i.e., $m=n=N$, the Birkhoff--von Neumann theorem ensures that an optimal transport plan $\mathbf{P}^{\ast}=\min_{\mathbf{P}\in\Pi(\mathbf{\mu},\mathbf{\nu})}
\sum_{i=1}^{m}\!\sum_{j=1}^{n}\mathbf{P}_{ij}\mathbf{C}_{ij}$ is a permutation matrix~\cite[Proposition~2.1]{peyre2019computational}. Consequently, in this setting, the discrete OT problem is equivalent to a minimum-cost bipartite matching problem. This equivalence provides the theoretical basis for the subsequent analysis of the relationship between pseudo-label shift and the detection error rate.

\subsection{Average Confidence and Wasserstein Distance}
For a deployment sample $\mathbf{x}^{(i)}$, the pseudo-label $\hat{\mathbf{y}}^{(i)}$ generated by detector $f$ is
\begin{equation}
	\hat{\mathbf{y}}^{(i)}_k=
	\begin{cases}
		1, & k=\mathrm{dec}\left(f\left( \mathbf{x}^{(i)}\right) \right) ,\\
		0, & \mathrm{otherwise}.
	\end{cases}
	\label{eq:111}
\end{equation}
Let ${f}_{\#}P_{\mathrm{D}}(\hat{\mathbf{y}})$ denote the distribution of pseudo-labels generated by $f$ on the deployment echo dataset $\mathcal{D}_{\mathrm{dep}} \sim P_\mathrm{D}(\mathbf{x})$, and let $P_\mathrm{D}(\mathbf{y})$ denote the corresponding ground-truth label distribution. Here, ${f}_{\#}P_{\mathrm{D}}$ denotes the pushforward distribution induced by applying detector $f$ to $P_{\mathrm{D}}(\mathbf{x})$. Similarly, the distribution of probability vectors $\mathbf{p}=f\left( \mathbf{x}\right)$ generated by $f$ on $P_\mathrm{D}(\mathbf{x})$ is denoted by ${f}_{\#}P_{\mathrm{D}}(\mathbf{p})$. The distribution of label vectors $\mathbf{y}$ in the validation set $\mathcal{D}_{\mathrm{val}}$ is denoted by $P_\mathrm{T}(\mathbf{y})$. Based on these definitions, Proposition~\ref{proposition1} establishes the equivalence between the AC-estimated error rate ${\mathcal{E}}_{\mathrm{AC}}$ and a Wasserstein distance.

\begin{figure*}[t]
	\centering
	\includegraphics[width=1.0\textwidth, height=0.29\textwidth]{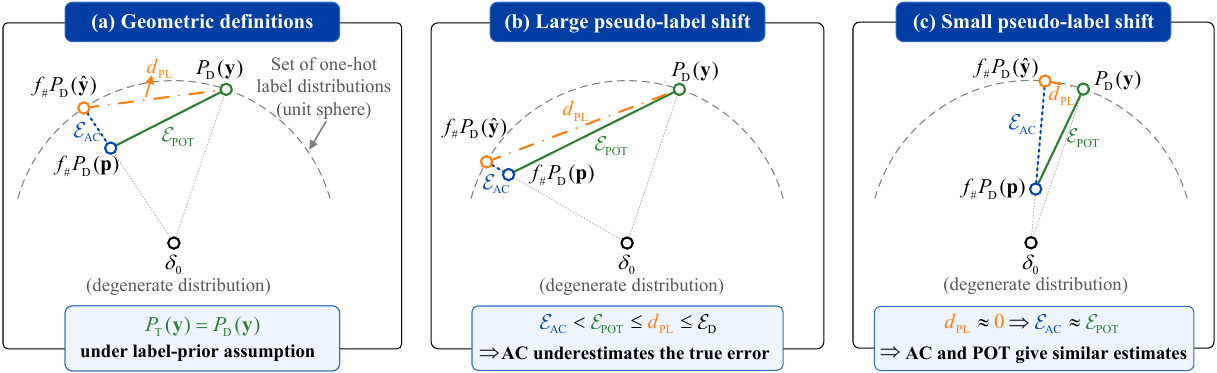}
	\vspace{-0.4cm}
	\caption{Geometric relationships among AC, POT, and pseudo-label shift in Wasserstein space. The blue segment represents the error rate estimated by AC, namely, the Wasserstein distance between the detector's output probability distribution ${f}_{\#}P_{\mathrm{D}}(\mathbf{p})$ and its pseudo-label distribution ${f}_{\#}P_{\mathrm{D}}(\hat{\mathbf{y}})$. Under the assumption $P_{\mathrm{T}}(\mathbf{y}) = P_{\mathrm{D}}(\mathbf{y})$, the green segment represents the POT error-rate estimate, namely, the Wasserstein distance between ${f}_{\#}P_{\mathrm{D}}(\mathbf{p})$ and the ground-truth label distribution $P_{\mathrm{D}}(\mathbf{y})$. The orange segment represents the degree of pseudo-label shift $d_{\mathrm{PL}}$, i.e., the distributional discrepancy between ${f}_{\#}P_{\mathrm{D}}(\hat{\mathbf{y}})$ and $P_{\mathrm{D}}(\mathbf{y})$. (a) AC projects ${f}_{\#}P_{\mathrm{D}}(\mathbf{p})$ onto the pseudo-label distribution induced by the detector itself and therefore tends to underestimate the true deployment error rate. (b) Under severe pseudo-label shift, ${\mathcal{E}}_{\mathrm{AC}}<d_{\mathrm{PL}}\le \mathcal{E}_{\mathrm{D}}$, indicating that AC underestimates the error rate, whereas POT provides a more reliable estimate by replacing the detector-induced pseudo-label distribution with the ground-truth label distribution as the reference. (c) When pseudo-label shift is small, AC and POT yield similar and relatively accurate estimates.}
	\label{fig:AC_Wasserstein}
\end{figure*}

\begin{proposition}\label{proposition1}
	Let $c_\infty(\mathbf{p},\mathbf{y})=\left\|\mathbf{p}-\mathbf{y}\right\|_\infty$ denote the transport cost between probability vector $\mathbf{p}\in\Delta^{K-1}$ output by detector $f$ and label vector $\mathbf{y}$, where $\left\|\cdot \right\|_\infty$ is the maximum norm. Let $W_{\infty}(\cdot,\cdot)$ denote the Wasserstein distance induced by this cost function. Then
	\begin{equation}
		{\mathcal{E}}_{\mathrm{AC}}=W_\infty\left( f_{\#}P_{\mathrm{D}}\left( \mathbf{p}\right) ,\ {f}_{\#}P_{\mathrm{D}}\left(\hat{\mathbf{y}}\right) \right).
	\end{equation}
\end{proposition}
\begin{remark}
	Fig.~\ref{fig:AC_Wasserstein} provides an intuitive geometric interpretation of Proposition~\ref{proposition1}. In the figure, $\delta_0$ denotes the degenerate distribution whose entire probability mass is concentrated at the zero vector and can be viewed as the origin of the distribution space. All one-hot label distributions, including $P_\mathrm{D}(\mathbf{y})$ and ${f}_{\#}P_{\mathrm{D}}(\hat{\mathbf{y}})$, lie on the unit sphere centered at $\delta_0$, whereas the detector's output probability-vector distribution ${f}_{\#}P_{\mathrm{D}}(\mathbf{p})$ lies inside this sphere. Proposition~\ref{proposition1} shows that the detector's deployment error-rate estimate obtained using AC, ${\mathcal{E}}_{\mathrm{AC}}$, is the distance from ${f}_{\#}P_{\mathrm{D}}(\mathbf{p})$ to the corresponding projection point on the unit sphere. This projection point is the pseudo-label distribution ${f}_{\#}P_{\mathrm{D}}(\hat{\mathbf{y}})$ induced by the detector outputs.
\end{remark}

The following corollary follows directly from Proposition~\ref{proposition1}.
\begin{corollary}\label{corollary1}
	Let $P(\mathbf{y})$ be any probability distribution defined over the set of one-hot label vectors. Then
	\begin{equation}
		W_\infty\!\left( {f}_{\#}P_{\mathrm{D}}(\mathbf{p}), P(\mathbf{y})\right)  \!\ge\! W_\infty\!\left( {f}_{\#}P_{\mathrm{D}}(\mathbf{p}), {f}_{\#}P_{\mathrm{D}}\!\left( \hat{\mathbf{y}}\right) \right)\!.
	\end{equation}
\end{corollary}
\begin{remark}
	Corollary~\ref{corollary1} shows that, among all possible reference label distributions, the pseudo-label distribution induced by the detector itself has the smallest Wasserstein distance to the detector's output probability distribution. AC therefore uses the reference label distribution that minimizes the transport cost and estimates the error rate using this minimum distance, thereby underestimating the detection error rate. This result provides a geometric explanation for the tendency of AC to overestimate deployment accuracy observed in previous studies~\cite{balasubramanian2011unsupervised, nguyen2026reliably, 9238468, chen2022estimating}. The subsequent analysis shows that this estimation bias becomes more pronounced when the detector exhibits substantial pseudo-label shift under the deployment distribution.
\end{remark}

\subsection{Pseudo-Label Shift and Estimation Error}\label{sec:Pseudo-Label Shift}
Based on OT theory, pseudo-label shift is quantified by the Wasserstein distance between the pseudo-label distribution ${f}_{\#}P_{\mathrm{D}}\left( \hat{\mathbf{y}}\right)$ and the ground-truth label distribution $P_\mathrm{D}(\mathbf{y})$:
\begin{equation}
	d_{\mathrm{PL}} = W_\infty({f}_{\#}P_{\mathrm{D}}\left( \hat{\mathbf{y}}\right), P_\mathrm{D}(\mathbf{y})).
\end{equation}
This quantity characterizes the distributional discrepancy between the pseudo-labels and the ground-truth labels. When
$d_{\mathrm{PL}}$ is large, the class composition output by the detector has deviated substantially from the true class proportions in the current deployment scenario. In Fig.~\ref{fig:AC_Wasserstein}, $d_{\mathrm{PL}}$ is represented by the length of the orange segment. Proposition~\ref{proposition2} establishes the relationship between pseudo-label shift and the detector's true error rate $\mathcal{E}_{\mathrm{D}}$.

\begin{proposition}\label{proposition2}
	The pseudo-label shift $d_{\mathrm{PL}}$ is a lower bound on the detector's true deployment error rate, i.e.,
	\begin{equation}
		W_\infty\left({f}_{\#}P_{\mathrm{D}}\left( \hat{\mathbf{y}}\right), P_\mathrm{D}(\mathbf{y})\right)
		\le \mathcal{E}_{\mathrm{D}} \le 1 .
		\label{eq_proposition2}
	\end{equation}
\end{proposition}
\begin{remark}
	The detector's true error rate $\mathcal{E}_{\mathrm{D}}$ is the average mismatch cost between the predicted and ground-truth labels for corresponding samples, whereas the Wasserstein distance between ${f}_{\#}P_{\mathrm{D}}\left( \hat{\mathbf{y}}\right)$ and $P_\mathrm{D}(\mathbf{y})$ selects the minimum transport cost over all possible matching schemes. Its OT cost therefore cannot exceed the average matching cost induced by the actual sample correspondence, and hence
	$W_\infty\left({f}_{\#}P_{\mathrm{D}}\left( \hat{\mathbf{y}}\right), P_\mathrm{D}(\mathbf{y})\right)\le\mathcal{E}_{\mathrm{D}}$. A detailed proof is given in Appendix~\ref{proof_3}. Proposition~\ref{proposition2} shows that the shift between the pseudo-label and ground-truth label distributions provides a valid lower bound on the detection error rate. A reliable error-rate estimator should therefore, to the extent possible, not produce an estimate below $W_\infty\left({f}_{\#}P_{\mathrm{D}}\left( \hat{\mathbf{y}}\right), P_\mathrm{D}(\mathbf{y})\right)$.
\end{remark}

Propositions~\ref{proposition1} and~\ref{proposition2} further characterize when and why AC underestimates the true deployment error rate. As illustrated in Fig.~\ref{fig:AC_Wasserstein}(b), when $\mathcal{E}_{\mathrm{AC}}<d_{\mathrm{PL}}$, it follows from~\eqref{eq_proposition2} that ${\mathcal{E}}_{\mathrm{AC}}<d_{\mathrm{PL}}\le \mathcal{E}_{\mathrm{D}}$.
Therefore, AC necessarily underestimates the detector's true error rate. The fundamental reason is that AC measures only the distance between the detector's output probability distribution and its self-induced pseudo-label distribution and ignores the shift of the pseudo-label distribution relative to the ground-truth label distribution. When the detector exhibits systematic class-decision bias, its probabilistic outputs may remain highly consistent with pseudo-labels constructed from its own predictions. Thus, even when the predicted class structure has become substantially distorted, AC may still produce a low error-rate estimate. Moreover, the degree of underestimation tends to become more pronounced as the magnitude of pseudo-label shift increases.

\begin{figure}[t]
	\centering
	\includegraphics[width=0.48\textwidth, height=0.34\textwidth]{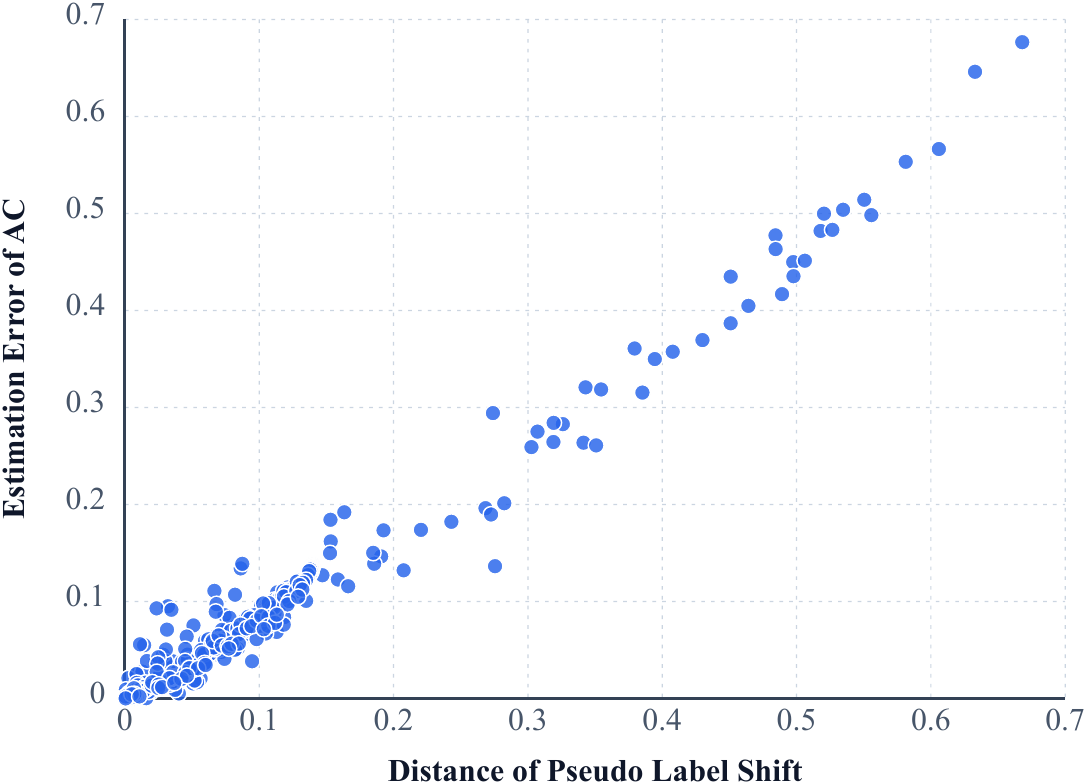}
	\caption{Relationship between pseudo-label shift and AC estimation error. The horizontal axis is the Wasserstein distance $d_{\mathrm{PL}}$ between the pseudo-label distribution ${f}_{\#}P_{\mathrm{D}}(\hat{\mathbf{y}})$ and the ground-truth label distribution $P_{\mathrm{D}}(\mathbf{y})$, and the vertical axis is the absolute error $\left|\mathcal{E}_{\mathrm{D}}-{\mathcal{E}}_{\mathrm{AC}}\right|$ between the AC-estimated and true error rates. Each point corresponds to a deployment dataset with a distribution mismatch.}
	\label{fig:pseudo_label_shift_estim_error_AC}
\end{figure}

To validate the preceding analysis, we further examine the relationship between pseudo-label shift $d_{\mathrm{PL}}$ and $\left|\mathcal{E}_{\mathrm{D}}-{\mathcal{E}}_{\mathrm{AC}}\right|$ using the radar target detection dataset, as shown in Fig.~\ref{fig:pseudo_label_shift_estim_error_AC}. The results show a pronounced positive correlation. AC generally provides relatively accurate performance estimates when pseudo-label shift is small. As the deployment distribution deviates further from the training distribution, the detector exhibits an increasingly pronounced pseudo-label shift, which induces systematic bias in its class decisions and, consequently, leads to a substantial increase in the AC estimation error. This result is consistent with the theoretical analysis and indicates that pseudo-label shift is a key factor affecting the reliability of AC-type methods.

\section{Methodology}\label{sec:method}
This section first presents POT for estimating the detection performance of a deployed detector and then introduces its robust variant, TPOT. Specifically, POT estimates the deployment error rate using the OT distance between the distribution of detector output probabilities and the label distribution of the echoes. TPOT further improves robustness against anomalous transport costs by adaptively thresholding the transport cost associated with each echo sample.

\subsection{POT}
As established in Section~\ref{sec:Pseudo-Label Shift}, the error rate estimate produced by AC equals the Wasserstein distance between the detector's output probability distribution $f_{\#}P_{\mathrm{D}}(\mathbf{p})$ and the detector-induced pseudo-label distribution $f_{\#}P_{\mathrm{D}}(\hat{\mathbf{y}})$. Because the pseudo-label distribution is generated entirely from the detector's own outputs, AC tends to underestimate the true deployment error rate when the detector exhibits substantial pseudo-label shift under the deployment distribution.

To address this limitation, POT replaces the detector-induced pseudo-label distribution with the ground-truth label distribution as the reference distribution for estimating the deployment error rate:
\begin{equation}
	{\mathcal{E}}_{\mathrm{POT}} = W_\infty\left( f_{\#}P_{\mathrm{D}}\left( \mathbf{p}\right), P_{\mathrm{D}}\left({\mathbf{y}}\right)\right).
\end{equation}
This distance quantifies the deviation of the detector output-probability distribution from the ground-truth label distribution of the deployment echoes, rather than from the detector's own predictions. Consequently, ${\mathcal{E}}_{\mathrm{POT}}$ does not directly depend on the detector-induced pseudo-label distribution ${f}_{\#}P_{\mathrm{D}}\left(\hat{\mathbf{y}}\right)$ and is therefore less susceptible to the estimation bias associated with pseudo-label shift. In Fig.~\ref{fig:AC_Wasserstein}, this distance is represented by the length of the green segment. 

However, the ground-truth labels associated with the echoes in $\mathcal{D}_{\mathrm{dep}}$ are unobserved, and therefore $W_\infty\left( f_{\#}P_{\mathrm{D}}\left( \mathbf{p}\right), P_{\mathrm{D}}\left({\mathbf{y}}\right)\right)$ cannot be computed directly. In the radar detection scenarios considered here, the set of target classes is assumed to remain unchanged between training and deployment, and the class-prior proportions are assumed not to vary substantially. Performance degradation is instead assumed to arise primarily from changes in the clutter distribution. Consequently, we have $P_{\mathrm{T}}\left({\mathbf{y}}\right) = P_{\mathrm{D}}\left({\mathbf{y}}\right)$, where $P_{\mathrm{T}}\left({\mathbf{y}}\right)$ denotes the training-stage label distribution. Accordingly, the POT estimator of the detector's error rate under the deployment distribution is defined as
\begin{equation} 
	{\mathcal{E}}_{\mathrm{POT}} = W_\infty\left( f_{\#}P_{\mathrm{D}}\left( \mathbf{p}\right), P_{\mathrm{T}}\left({\mathbf{y}}\right)\right) ,
	\label{eq_POT}
\end{equation}
where $f_{\#}P_{\mathrm{D}}(\mathbf{p})$ is constructed from the probability vectors output by the detector on deployment data $\mathcal{D}_{\mathrm{dep}}$, and $P_{\mathrm{T}}(\mathbf{y})$ is constructed from label statistics of the training-stage validation set $\mathcal{D}_{\mathrm{val}}$.

Combining Corollary~\ref{corollary1} with~\eqref{eq_POT} yields ${\mathcal{E}}_{\mathrm{POT}} \ge {\mathcal{E}}_{\mathrm{AC}}$. Thus, the error rate estimated by POT is theoretically no smaller than that estimated by AC. Because AC tends to underestimate the true deployment error rate of the detector, ${\mathcal{E}}_{\mathrm{POT}}$ can generally provide a more accurate performance estimate. Moreover, The deployment error rate estimated by AC is not lower-bounded by any positive function of the pseudo-label shift. Consequently, under substantial pseudo-label shift, AC may severely underestimate the true deployment error rate and yield an estimate close to zero, resulting in a large $\left|\mathcal{E}_{\mathrm{D}}-{\mathcal{E}}_{\mathrm{AC}}\right|$. By contrast, POT satisfies the following guarantee.
\begin{proposition}\label{proposition3}
	Under $P_{\mathrm{T}}\left({\mathbf{y}}\right) = P_{\mathrm{D}}\left({\mathbf{y}}\right)$, the error rate estimated by POT satisfies
	\begin{equation} 
		{\mathcal{E}}_{\mathrm{POT}} \ge \frac{1}{2} W_\infty\left( {f}_{\#}P_{\mathrm{D}}\left(\hat{\mathbf{y}}\right), P_{\mathrm{D}}\left({\mathbf{y}}\right)\right). 
	\end{equation}
\end{proposition}
\begin{remark}
	Proposition~\ref{proposition3} establishes a lower-bound relationship between the POT error-rate estimate and pseudo-label-distribution shift. As the pseudo-label shift increases, the corresponding lower bound on the POT estimate also increases. Consequently, when the pseudo-label distribution deviates substantially from the ground-truth label distribution, this lower bound prevents POT from yielding a spuriously near-zero error-rate estimate, as may occur with AC, and the POT estimate can therefore be closer to the true deployment error rate than the AC estimate. The proof of Proposition~\ref{proposition3} is provided in Appendix~\ref{proof_4}, and the tightness of the resulting lower bound is established in Appendix~\ref{proof_5}.
\end{remark}

\subsection{TPOT}
The procedure for estimating the detector's deployment error rate using POT consists of three stages. First, a transport cost matrix is constructed by computing the pairwise distances between the detector's output probability vectors and the reference label vectors. Second, an OT plan ${\mathbf{P}}^{\ast}$ is obtained by minimizing the total transport cost subject to the prescribed marginal constraints. Third, the transport costs selected by $\mathbf{P}^{\ast}$ are averaged to compute the Wasserstein distance $ W_\infty\left( f_{\#}P_{\mathrm{D}}\left( \mathbf{p}\right), P_{\mathrm{T}}\left({\mathbf{y}}\right)\right)$, thereby yielding the estimated deployment error rate ${\mathcal{E}}_{\mathrm{POT}}$.

Because the POT estimate is based on the mean of the OT costs associated with the radar echo samples, it can be sensitive to anomalously large values~\cite{rousseeuw2011robust,garg2022leveraging}. In the presence of strong clutter, at low signal-to-clutter ratios, or under anomalous sensing conditions, some detector outputs may deviate substantially from their typical patterns and yield excessively large transport costs. These anomalous costs can disproportionately affect the mean transport cost used by POT, thereby degrading estimation accuracy and stability.

To mitigate the sensitivity of POT to anomalously large transport costs, we propose TPOT, a thresholded variant of POT. Rather than averaging all OT costs selected by $\mathbf{P}^{\ast}$, TPOT first learns a threshold $t^{\ast}$ from the validation set based on the detector's error rate on the validation set. Then, $t^{\ast}$ is applied to all OT costs selected by the transport plan $\mathbf{P}^{\ast}$ obtained using POT on deployment data $\mathcal{D}_{\mathrm{dep}}$. The proportion of these costs that exceed $t^{\ast}$ is used as the estimate of the deployment error rate. After thresholding, each radar echo sample contributes only a binary value, either $0$ or $1$, to the final estimate. Once an OT cost exceeds $t^{\ast}$, any further increase in its magnitude does not change the contribution of the corresponding radar echo sample, thereby limiting the influence of anomalously large transport costs. The threshold calibration and deployment error-rate estimation procedures are detailed below.

First, the OT plan between the detector's output probability distribution and the ground-truth label distribution is computed on the validation set $\mathcal{D}_{\mathrm{val}}$:
\begin{equation}
	{\mathbf{P}}^{\ast}_{\mathrm{val}} = \arg\min_{\mathbf{P} \in \Pi(f_\# {P}_{\mathrm{T}}(\mathbf{p}), {P}_{\mathrm{T}}(\mathbf{y}))} \langle \mathbf{P}, \mathbf{C} \rangle,
\end{equation}
where $\mathbf{C}$ is the cost matrix computed using the $L_\infty$ distance. For the validation set, the numbers of detector-output probability vectors and radar-echo label vectors are equal, both being $N_{\mathrm{val}}$. The resulting discrete OT problem therefore reduces to a linear assignment problem, and the OT plan ${\mathbf{P}}_{\mathrm{val}}^{\ast}$ is a permutation matrix~\cite{peyre2019computational}. Consequently, each detector-output probability vector is matched to exactly one reference label vector, yielding one OT cost for each radar echo sample in the validation set. 

The TPOT threshold $t^{\ast}\in[0,1]$ is then calibrated on the validation set such that the proportion of radar echo samples in the validation set whose matched OT costs are no smaller than $t^{\ast}$ is as close as possible to the detector's actual error rate on the validation set:
\begin{equation}
	t^{\ast} \!=\! \arg\min_{t\in[0,1]} \!\left| \frac{1}{\left|\mathbf{P}^{\ast}_{\mathrm{val}} \right|}\! \sum_{\mathbf{C}_{ij}\in \mathbf{C}} \!\!\mathbb{I}\left( \mathbf{P}^{\ast}_{\mathrm{val}}\mathbf{C}_{ij} \ge t\right) 
	- \mathcal{E}_{\mathrm{val}} \right|, 
\end{equation}
where $\mathbf{C}_{ij}$ is the transport cost between the detector output $\mathbf{p}^{(i)}$ for the $i$-th sample $\mathbf{x}^{(i)}$ and the label $\mathbf{y}^{(j)}$ of the $j$-th sample. By probabilistically calibrating $t^{\ast}$ using the detector's true error rate, TPOT establishes a statistical correspondence between the calibrated threshold and the detector's actual detection-error events.

During performance estimation, the OT plan between the detector outputs on the deployment data $\mathcal{D}_{\mathrm{dep}}$ and the label distribution ${\mathbf{P}}_{\mathrm{T}}(\mathbf{y})$ is computed as:
\begin{equation}
	\mathbf{P}^{\ast}_{\mathrm{dep}} = \arg\min_{\mathbf{P} \in \Pi({f}_\# {P}_{\mathrm{D}}(\mathbf{p}), {\mathbf{P}}_{\mathrm{T}}(\mathbf{y}))} \langle \mathbf{P}, \mathbf{C} \rangle.
\end{equation} 
The threshold $t^{\ast}$ is then applied to the transport costs selected by $\mathbf{P}^{\ast}_{\mathrm{dep}}$. Each selected cost is converted into a binary indicator that equals $1$ if the cost is no smaller than $t^{\ast}$ and $0$ otherwise. Finally, the deployment error rate is estimated by averaging the resulting binary indicators:
\begin{equation} 
	{\mathcal{E}}_{\mathrm{TPOT}} = \frac{1}{|\mathbf{P}^{\ast}_{\mathrm{dep}} | } \left|\left\{\mathbf{C}_{ij} \ge t^{\ast} \mid \mathbf{P}^{\ast}_{\mathrm{dep}, ij} = 1\right\}\right|. 
\end{equation}

\begin{table}[t]
	\renewcommand{\arraystretch}{1.3}
	\caption{Radar parameters of the MTDSP dataset}
	\label{Radar parameters}
	\centering
	\setlength{\tabcolsep}{1.0mm}{ 
		\begin{tabular}{cccc}
			\bottomrule
			\cmidrule{1-4} 
			{\textbf{Bandwidth}} & {\textbf{Power}} & {\textbf{Resolution}} & {\textbf{Signal Type}} \\
			\cmidrule{1-4} 
			25 MHz &  100 W &  6m& LFM\\
			\bottomrule
			\cmidrule{1-4}
			{\textbf{Carrier Frequency}} & {\textbf{PRF}} & {\textbf{Sampling Frequency}} & {\textbf{Pulse Width}} \\
			\cmidrule{1-4} 
			9.4 GHz & 2000 Hz & 60 MHz & 8 us \\
			\bottomrule
			\cmidrule{1-4}
	\end{tabular}}
\end{table}

\begin{figure}[t]
	\centering
	\includegraphics[width=0.48\textwidth, height=0.24\textwidth]{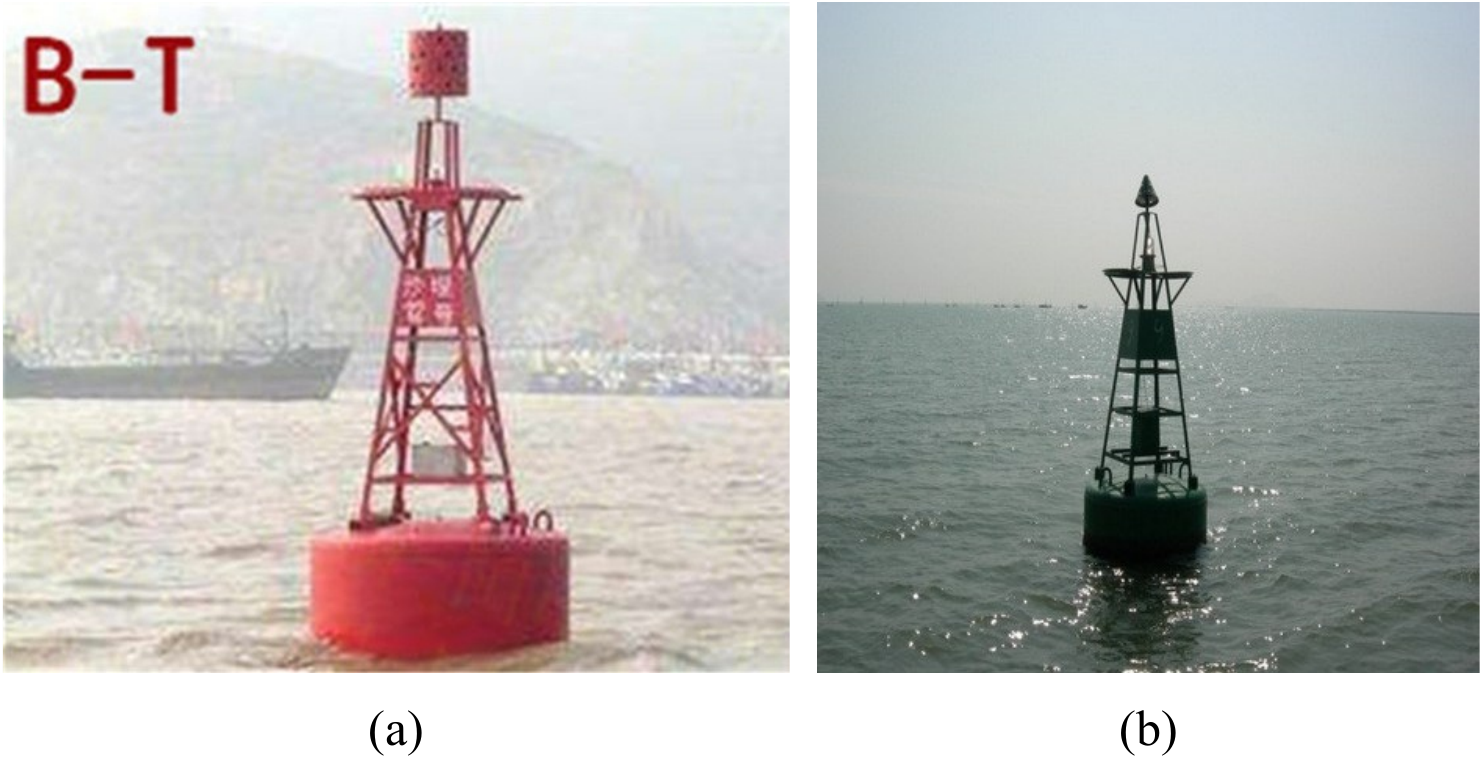}
	\caption{Observed targets. (a) A passing cargo ship and a steel buoy in the near-range region. (b) A steel buoy in the far-range region.}
	\label{fig_target}
\end{figure}

\section{Experiments}\label{sec:experiments}
This section first describes the experimental setup, including the dataset, detector architecture, training and implementation details, and evaluation metrics. The proposed POT and TPOT methods are then compared with representative label-free performance estimation methods to evaluate their effectiveness and robustness.

\subsection{Dataset Description}
The experiments use the Maritime Target Data Sharing Program (MTDSP) database constructed by Naval Aviation University~\cite{jian2023sea}. The data in the database were was collected using an SPPR50P X-band solid-state coherent radar deployed in Yantai, Shandong Province, China; the radar operating parameters are listed in Table~\ref{Radar parameters}. The observed objects include two offshore steel buoys and cargo ships passing through the observation area, as shown in Fig.~\ref{fig_target}. Data were collected from 14:00 on November 12, 2022, to 08:00 on November 15, 2022, under sea states 2--5.

\begin{figure*}[t]
	\centering
	\includegraphics[width=1.0\textwidth, height=0.27\textwidth]{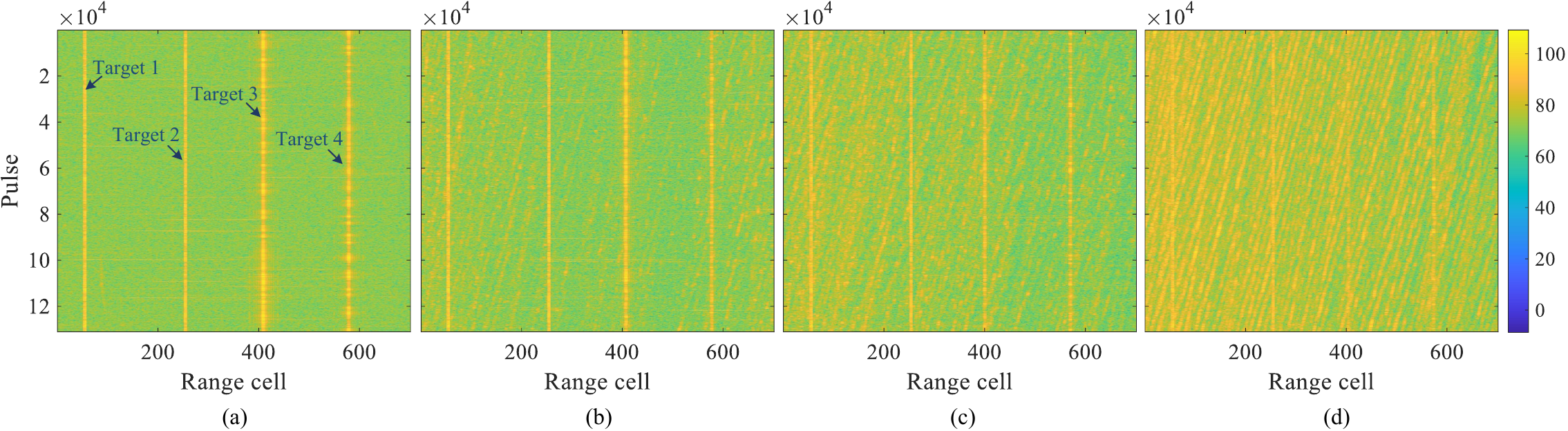}
	\vspace{-0.5cm}
	\caption{MTDSP dataset. (a)--(d) Radar echoes under sea states 2--5. Target 1 is a passing cargo ship, Target 2 is an injected simulated target, and Targets 3 and 4 are offshore steel buoys. Sea-clutter interference increases with sea state.}
	\label{fig:MTDSP}
\end{figure*}

To increase target-class diversity, echoes corresponding to an additional simulated target class were injected into the measured data. The experimental data therefore contain four target classes, and representative echoes under different sea states are shown in Fig.~\ref{fig:MTDSP}. The data were acquired in groups at hourly intervals, with each group comprising 65.5 s of continuous observations. In total, 71 groups were obtained, each containing 131,072 pulses. Because sea clutter is strongly nonstationary and time-varying, the echo statistics differ to varying degrees among data groups. The database therefore provides representative distribution-shift scenarios based on measured radar data for evaluating label-free performance estimation methods in complex environments.

\begin{figure}[t]
	\centering
	\includegraphics[width=0.48\textwidth, height=0.59\textwidth]{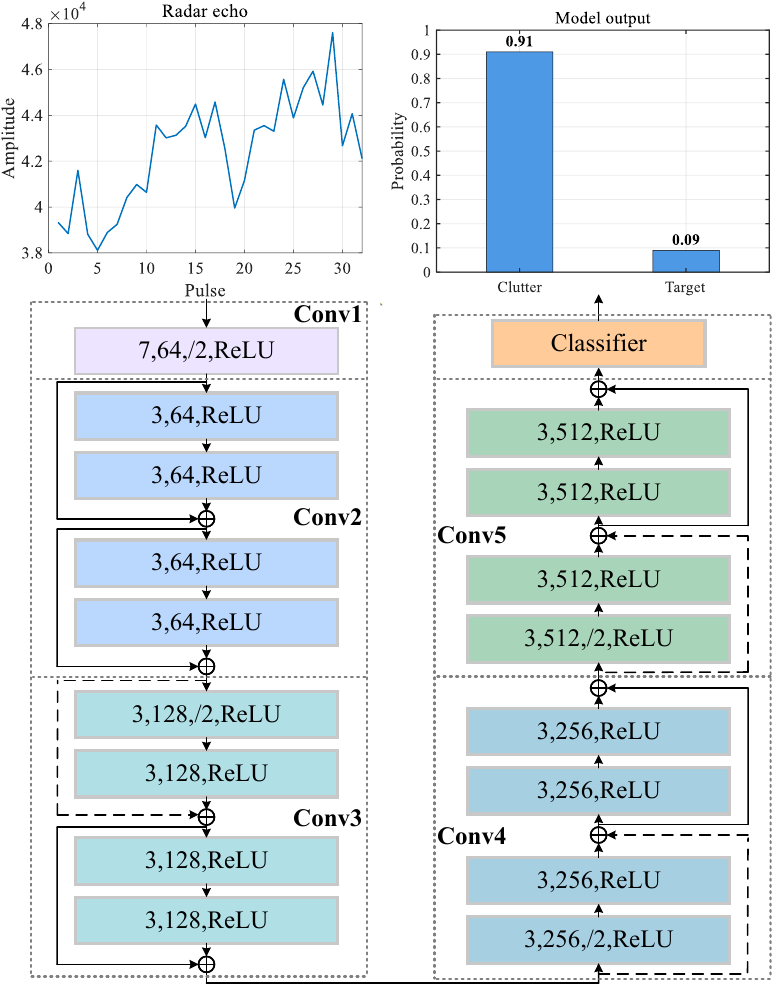}
	\caption{Detector architecture. In the figure, $7,64,/2,\text{ReLU}$ indicates that the corresponding convolutional module uses a kernel size of 7, 64 output channels, a stride of 2, and a ReLU activation function. Each convolutional layer is followed by batch normalization. The radar-signal magnitude image is included only to illustrate the echo form for architectural visualization and is not the actual network input. }
	\label{resnet18}
\end{figure}

\subsection{Implementation Details}
A ResNet is used as the backbone of the radar target detector, and the number of input pulses for each CUT is set to $P=64$. Residual connections transmit information across layers and help alleviate vanishing gradients during deep-network training~\cite{9143512}. The standard ResNet architecture is modified to accommodate the characteristics of radar echo data, as shown in Fig.~\ref{resnet18}. Specifically, the max-pooling layer after the Conv1 module is removed to reduce the loss of fine-grained echo information caused by early downsampling. For complex-valued radar echoes, the real and imaginary parts are extracted and supplied to the detector as two independent input channels. The network parameters are initialized using Kaiming initialization, stochastic gradient descent (SGD) is used for optimization, and the initial learning rate is set to $0.001$.

To evaluate detector performance training--deployment distribution mismatch, a cross-group testing strategy is adopted. In each experiment, one of the 71 MTDSP data groups is selected as source-domain data for detector training and model selection, and each of the remaining 70 groups serves as an unlabeled deployment scenario. Because sea-clutter statistics vary dynamically with observation time and sea state, these deployment groups exhibit different degrees of distribution shift relative to the source-domain data. For each deployment scenario, a label-free performance estimation method first estimates the detector error rate. Ground-truth labels, available only for experimental evaluation, are then used to compute the actual error rate and assess estimation accuracy.

Each source-domain data group is partitioned into training, validation, and source-domain test sets in a 4:2:4 ratio. The training set is used to learn detector parameters, the validation set is used for model selection and parameter determination for the performance estimation methods, and the test set is used to evaluate source-domain detection performance. To reduce sensitivity to source-domain selection, two data groups are selected from each sea-state level, and a separate detector is trained on each selected group. Because four sea-state levels are considered, the experiments yield $2\times4\times70$ performance estimation results. The source-domain data groups used to train the detectors are listed in Table~\ref{table_train_data}. All experiments were conducted on a high-performance computing platform equipped with two Intel(R) Xeon(R) Platinum 8380 CPUs, 1024 GB of memory, and eight NVIDIA A40 GPUs. The operating system was Ubuntu 22.04.4 LTS, and the deep learning framework was PyTorch 2.4.0.

\begin{table}[t]
	\renewcommand{\arraystretch}{1.3}
	\caption{MTDSP dataset used to train the detectors}
	\label{table_train_data}
	\centering
	\setlength{\tabcolsep}{1.0mm}{ 
		\begin{tabular}{ccc}
			\bottomrule
			\cmidrule{1-3} 
			{\bf{Sea State}} & \multicolumn{2}{c}{{\bf{Data Name}}} \\
			\bottomrule
			\cmidrule{1-3}
			2 & 20221115040022\_stare\_HH & 20221115070036\_stare\_HH \\ 
			3 & 20221114030013\_stare\_HH & 20221114070031\_stare\_HH \\
			4 & 20221112150043\_stare\_HH & 20221113230100\_stare\_HH \\ 
			5 & 20221112205109\_stare\_HH & 20221113030022\_stare\_HH \\
			\bottomrule
			\cmidrule{1-3}
	\end{tabular}}
\end{table}

\subsection{Evaluation Metrics}
Although the proposed performance estimation methods do not require deployment-data labels, labeled deployment samples are used during experimental evaluation to compute the detector's true error rate and quantitatively assess the estimation accuracy of different methods. For the $m$-th deployment scenario, the true detector error rate is defined as
\begin{equation}
	\mathcal{E}_{\mathrm{D},m}^{\mathrm{True}} = \mathbb{E}_{(\mathbf{x},\mathbf{y})\sim \mathcal{D}_{\mathrm{dep}}}
	\left[ \mathbb{I}\left( \mathrm{dec}(f\left(\mathbf{x}\right)) \neq \mathrm{dec}(\mathbf{y}) \right) \right],
\end{equation}
Let ${\mathcal{E}}_{\mathrm{D},m}$ be the error-rate estimate produced by a given method. The absolute estimation error in this scenario is defined as
\begin{equation}
	\epsilon_{m}=\left|\mathcal{E}_{\mathrm{D},m}^{\mathrm{True}}-{\mathcal{E}}_{\mathrm{D},m}\right|.
\end{equation}

Averaging the absolute errors over all scenarios gives the conventional mean absolute error (MAE). In practice, however, the true deployment error rates across deployment scenarios may be highly imbalanced. When most scenarios have low true error rates, MAE can be dominated by low-error scenarios and may not adequately reflect estimation performance in high-error scenarios.

To mitigate evaluation bias caused by an imbalanced true-error-rate distribution, balanced mean absolute error (BMAE) is used as the primary evaluation metric. All deployment scenarios are first partitioned into $B$ error-rate intervals $\{I_b\}_{b=1}^{B}$ according to the true detector error rate $\mathcal{E}_{\mathrm{D},m}^{\mathrm{True}}$. For the $b$-th interval, let the set of scenario indices falling within the interval be
\begin{equation}
	\mathcal{B}_b = \left\{ m \mid \mathcal{E}_{\mathrm{D},m}^{\mathrm{True}} \in I_b \right\}.
\end{equation}
The MAE within this interval is
\begin{equation}
	\mathrm{MAE}_b = \frac{1}{|\mathcal{B}_b|} \sum_{m\in\mathcal{B}_b}
	\left| {\mathcal{E}}_{\mathrm{D},m} 	- \mathcal{E}_{\mathrm{D},m}^{\mathrm{True}} \right|.
\end{equation}
Finally, the $\mathrm{MAE}_b$ values of all nonempty error-rate intervals are averaged with equal weights to obtain BMAE:
\begin{equation}
	\mathrm{BMAE} = \frac{1}{B} \sum_{b} \mathrm{MAE}_b.
\end{equation}
A smaller BMAE indicates higher overall estimation accuracy across different error-rate intervals.

\subsection{Baselines}
The proposed POT and TPOT methods are compared with the following representative label-free performance estimation methods:

\textbullet~AC~\cite{ovadia2019can}: The estimator is given in~\eqref{eq:AC_Estimate}.

\textbullet~ATC-MC~\cite{garg2022leveraging}: The estimator is given in~\eqref{eq:ATC_Estimate}, with the maximum detector's output confidence used as the score function $s(\cdot)$.

\textbullet~ATC-NE~\cite{garg2022leveraging}: The estimator is computed as in ATC-MC, but the negative entropy of the detector's output probability distribution is used as $s(\cdot)$.

\textbullet~DoC~\cite{guillory2021predicting}: The detector error rate is estimated by correcting the validation error using the difference between the average confidence-based errors on the validation and deployment sets:
\begin{equation}
	\begin{aligned}
		&{\mathcal{E}}_{\text{DoC}} = \mathbb{E}_{(\mathbf{x},\mathbf{y})\sim \mathcal{D}_{\mathrm{val}}} \left[ \mathbb{I}\left[ \mathrm{dec}\left( f\left( \mathbf{x}\right) \right)  \!\neq\! \mathrm{dec}\left(\mathbf{y}\right)\right] \right] \\
		&+\! \mathbb{E}_{\mathbf{x} \!\sim\! \mathcal{D}_{\mathrm{dep}}} \!\!\left[ 1 \!-\! \max\!\left(  f\!\left(\mathbf{x}\right)\right)\right]  \!-\! \mathbb{E}_{\mathbf{x} \!\sim\! \mathcal{D}_{\mathrm{val}}}\!\! \left[ 1 \!-\! \max\!\left(  f\!\left(\mathbf{x}\right)\right)\right].\!\!
	\end{aligned}
\end{equation}

\textbullet~IM~\cite{liu2015classification}: The deployment error rate is estimated by reweighting the validation-set errors. The weights are computed from the ratio of the deployment-set and validation-set sample proportions within each score interval. Specifically, samples are partitioned into $B$ intervals $\{I_b\}_{b=1}^{B}$ according to the detector's output score $s(f\left(\mathbf{x}\right))$, and the ratio between the proportions of deployment and validation samples in the $b$-th interval is computed as
\begin{equation}
	w_b =
	\frac{
		\left|\{\mathbf{x}\in\mathcal{D}_{\mathrm{dep}} \mid s(f\left(\mathbf{x}\right))\in I_b\}\right|/N_{\mathrm{dep}}
	}{
		\left|\{\mathbf{x}\in\mathcal{D}_{\mathrm{val}} \mid s(f\left(\mathbf{x}\right))\in I_b\}\right|/N_{\mathrm{val}}
	}.
\end{equation}
The validation error is then reweighted using these weights to obtain the deployment error-rate estimate
\begin{equation}
	{\mathcal{E}}_{\mathrm{IM}} \!=\!\! \frac{1}{N_{\mathrm{val}}} \!\sum_{i=1}^{N_{\mathrm{val}}}
	\!w_{b(i)} \mathbb{I}\! \left(\! \mathrm{dec}\left( \!f\left(\! \mathbf{x}^{(i)}_{\mathrm{val}}\right)\! \right)  \!\neq\! \mathrm{dec}\!\left( \mathbf{y}^{(i)}_{\mathrm{val}}\right) \!\right),
\end{equation}
where $b_{(i)}$ denotes the score interval containing validation sample $\mathbf{x}^{(i)}_{\mathrm{val}}$.
	
\begin{figure}[t]
	\centering
	\includegraphics[width=0.48\textwidth, height=0.34\textwidth]{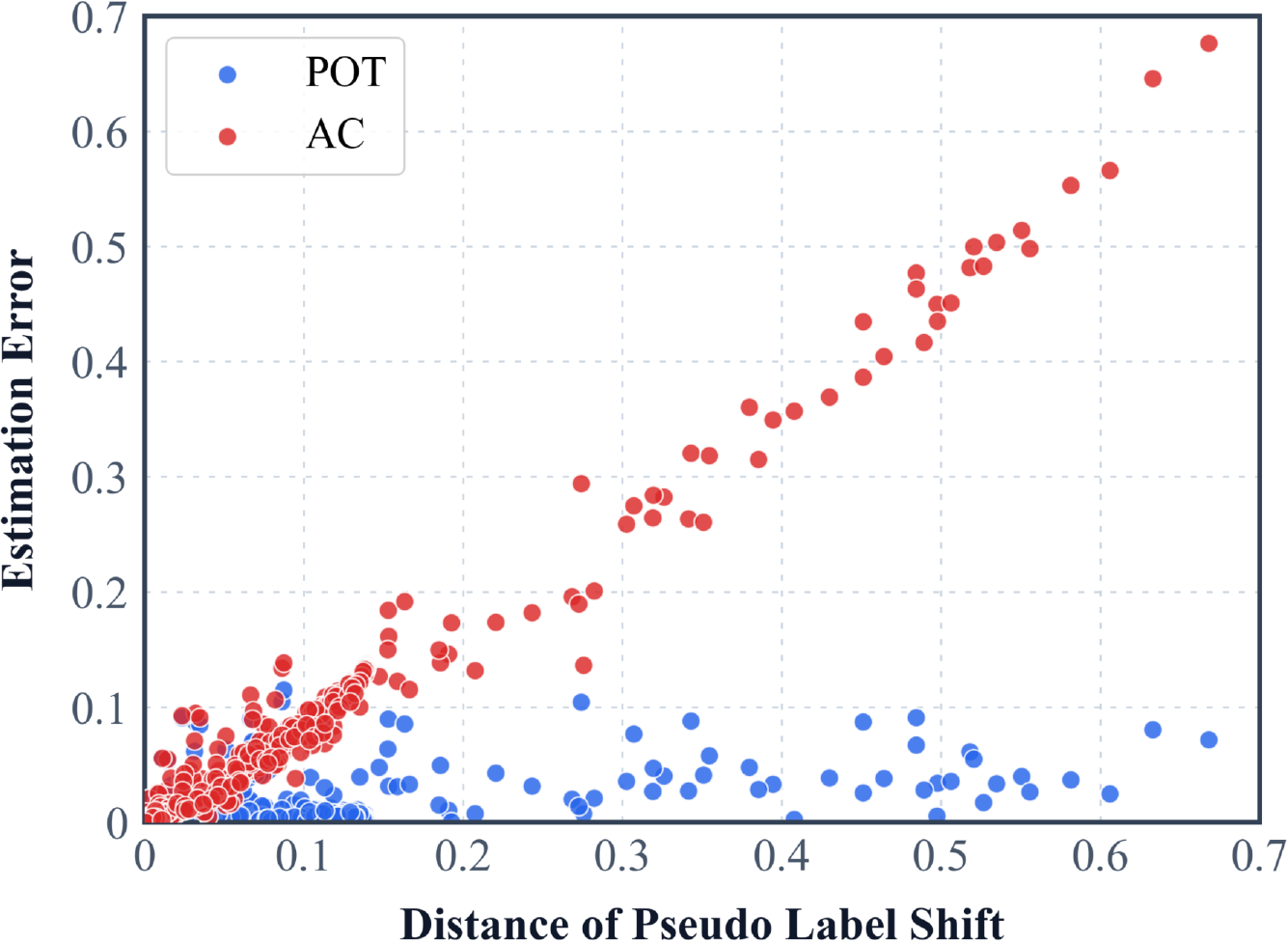}
	\caption{Relationship between pseudo-label shift and the absolute error in error-rate estimation for different methods.}
	\label{fig:pseudo_label_shift_estim_error_POT}
\end{figure}

\subsection{Experimental Results}
This section addresses two central questions in unlabeled deployment settings: whether the proposed methods accurately reflect the detector's true error rate and whether this estimation capability is maintained as pseudo-label shift, detector degradation, and the number of classes increase. We first examine the response of estimation error $\epsilon_{m}$ to pseudo-label shift, then compare scenario-level estimation accuracy, interval-wise MAE, and overall BMAE in binary and multiclass tasks, and finally evaluate the consistency between the empirical observations and the theoretical lower bounds.

\subsubsection{Estimation Behavior Under Pseudo-Label Shift}
Fig.~\ref{fig:pseudo_label_shift_estim_error_POT} shows the absolute estimation errors $\epsilon_{m}$ of AC and POT as functions of the pseudo-label shift distance $d_{\mathrm{PL}}$. The estimation error of AC generally increases with $d_{\mathrm{PL}}$, whereas the POT results do not exhibit a comparable upward trend over the examined range. Moreover, most POT estimation errors remain below $0.1$, indicating that POT is less sensitive to pseudo-label shift than AC.

This empirical observation demonstrates the effectiveness of using the training label distribution $P_{\mathrm{T}}(\mathbf{y})$, rather than the detector-induced pseudo-label distribution ${f}_{\#}P_{\mathrm{D}}(\hat{\mathbf{y}})$, as the reference distribution in POT. Because $P_{\mathrm{T}}(\mathbf{y})$ does not vary with the current pseudo-label distribution, the absolute POT estimation error, $| \mathcal{E}_{\mathrm{D}}^{\mathrm{True}} - {\mathcal{E}}_{\mathrm{POT}} |$,
does not exhibit a clear dependence on $d_{\mathrm{PL}}$ over the examined deployment scenarios. This result indicates that replacing the prediction-dependent pseudo-label distribution with the training label distribution as the reference enables POT to maintain stable error-rate estimation under increasing pseudo-label shift.

\begin{figure*}[t]
	\centering
	\includegraphics[width=0.99\textwidth, height=0.22\textwidth]{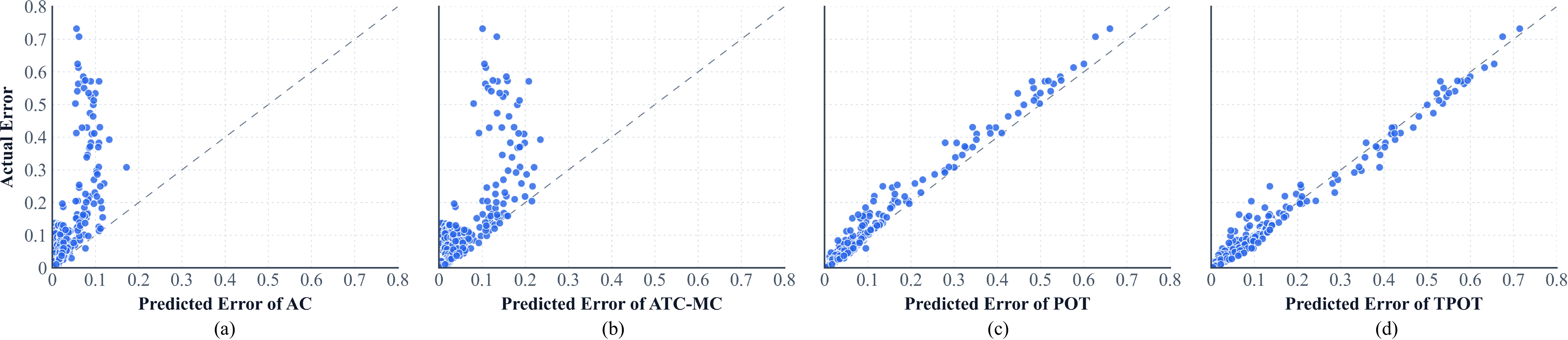}
	\caption{Error-rate estimates of different label-free performance estimation methods in the binary radar detection task. The horizontal and vertical axes show the estimated and true error rates, respectively. Each point corresponds to a distribution-mismatched deployment data group, and the dashed line indicates the ideal relationship $y=x$. AC~\cite{ovadia2019can} and ATC~\cite{garg2022leveraging} substantially underestimate the error rate in high-error scenarios, with points lying mainly above and to the left of the reference line. By contrast, the POT and TPOT points are closer to the reference line, indicating that the proposed methods more accurately reflect performance degradation under distribution shift.}
	\label{fig:ood_error_scatter_comp}
\end{figure*}

\subsubsection{Comparison of Estimation Accuracy}
To compare error-rate estimation accuracy across deployment scenarios, Fig.~\ref{fig:ood_error_scatter_comp} plots the error-rate estimates produced by AC, ATC-MC, POT, and TPOT against the corresponding true error rates in the binary radar target detection task. The dashed line $y=x$ represents perfect agreement between the estimated and true error rates.

As shown in Fig.~\ref{fig:ood_error_scatter_comp}(a), AC substantially underestimates the true deployment error rate in many deployment scenarios, particularly at high true error rates, where the points are concentrated above and to the left of the reference line $y=x$. This result indicates that, under distribution shifts caused by changing sea-clutter conditions, the detector may remain highly confident in incorrect predictions, and average confidence no longer reliably reflects the true accuracy. Fig.~\ref{fig:ood_error_scatter_comp}(b) shows that ATC-MC improves on AC to some extent after the confidence threshold is calibrated on the validation set, but substantial underestimation remains in the high-error region. ATC-MC depends on the relationship established on the validation set between the score threshold and error events. This relationship may drift when the deployment distribution changes substantially and induces pseudo-label shift. Consequently, ATC-MC still has difficulty accurately estimating the degree of detector degradation in high-error scenarios.

By contrast, the POT points in Fig.~\ref{fig:ood_error_scatter_comp}(c) lie substantially closer to the $y=x$ reference line, indicating more accurate estimation of the true detector error rate under distribution mismatch. This advantage arises from distribution-level comparison: rather than considering only the maximum confidence of individual samples or a fixed score threshold, POT uses OT theory to measure the overall deviation between the distribution of the detector's output probability vectors on the deployment data and the training label distribution. Thus, when the detector develops class-decision bias in complex sea clutter, POT can capture the mismatch through changes in the overall output distribution. The TPOT points in Fig.~\ref{fig:ood_error_scatter_comp}(d) are closest overall to the reference line and maintain good agreement with the corresponding true deployment error rates, even in high-error scenarios. This result indicates that adaptively calibrating a transport-cost threshold on the validation set and counting the proportion of transport events exceeding that threshold can reduce the influence of a small number of anomalous transport costs on the mean-based POT estimate, thereby improving robustness.

\begin{figure}[t]
	\centering
	\includegraphics[width=0.48\textwidth, height=0.36\textwidth]{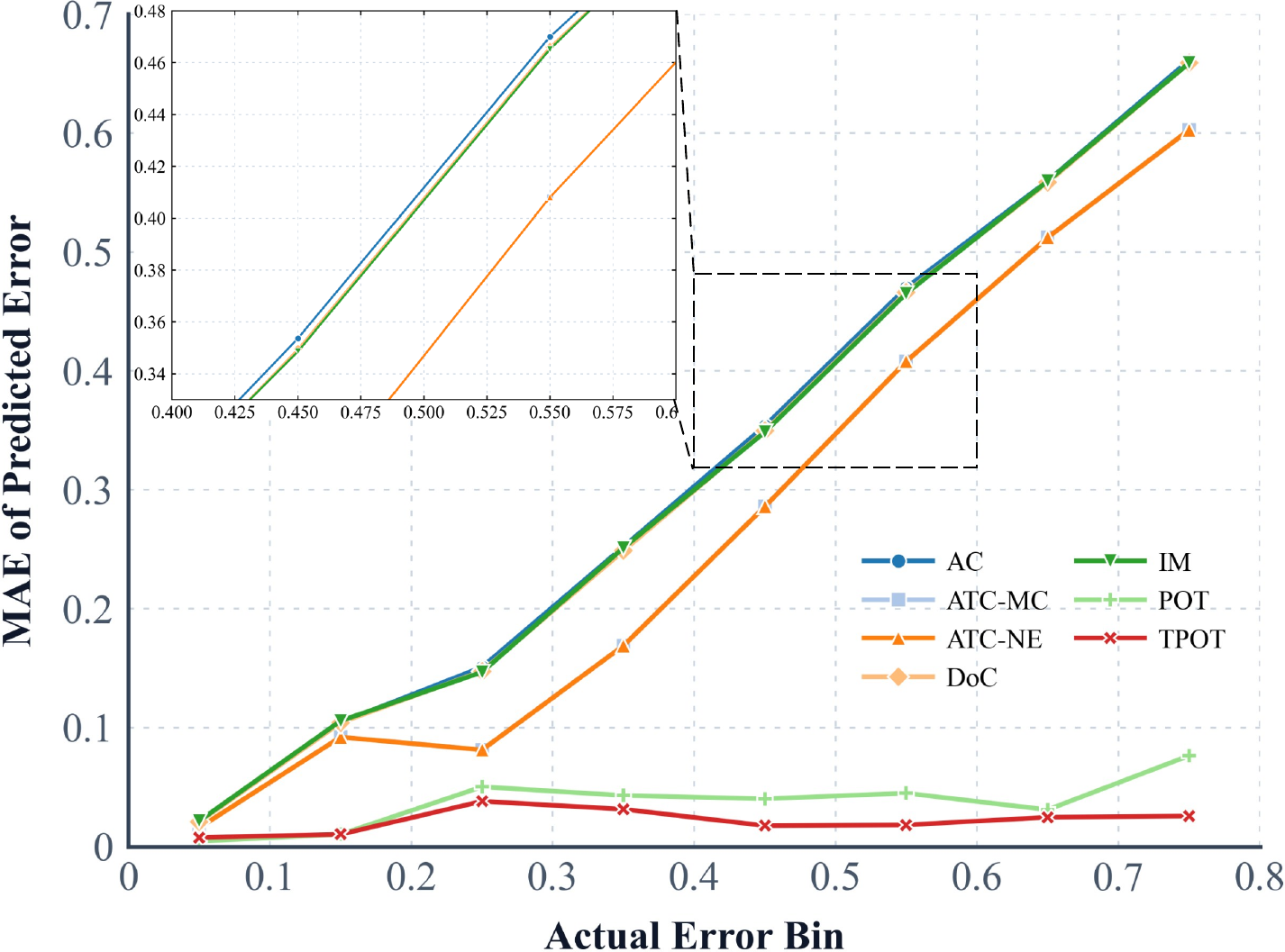}
	\caption{Mean absolute errors of different methods over true-error-rate intervals in the binary radar detection task. The horizontal axis denotes the true-error-rate interval, and the vertical axis denotes the mean estimation error within that interval.}
	\label{fig:ood_error_binned_mae_comparison}
\end{figure}

In addition to the scatter-plot analysis, we compute BMAE for quantitative evaluation. Fig.~\ref{fig:ood_error_binned_mae_comparison} shows the MAE curves over different true-error-rate intervals. The interval-wise MAE of AC, IM, DoC, and the ATC methods increases markedly with the true deployment error rate, indicating that these methods cannot provide reliable warning signals when the detector is severely degraded. In online radar systems, such high-error scenarios are precisely those in which model updating or adaptive training is most necessary; underestimating the true deployment error rate therefore creates substantial engineering risk. POT considerably reduces the estimation error in most error-rate intervals, and TPOT achieves the lowest or nearly lowest interval-wise errors. These results show that optimal-transport-based distribution matching improves not only overall estimation accuracy but also consistency across different degrees of performance degradation.

\subsubsection{Effect of the Number of Target Classes}
To examine the influence of the number of classes on label-free error-rate estimation, Table~\ref{loss_compar} reports the BMAE of each method for different values of $K$. As $K$ increases from 2 to 5, the estimation error of every method increases, indicating that multiclass radar detection substantially increases the difficulty of label-free performance estimation. With more classes, the detector's output probability space becomes higher dimensional, interclass confusion patterns become more complex, and class-prior proportions, interclass similarity, and pseudo-label shift may exhibit increasingly complex variations. A single confidence statistic is therefore less able to characterize performance degradation under the deployment distribution. Nevertheless, POT and TPOT clearly outperform the baselines for all tested values of $K$. For example, at $K=5$, the BMAE values of AC, IM, DoC, ATC-MC, and ATC-NE are 38.28\%, 36.03\%, 35.11\%, 31.82\%, and 31.41\%, respectively, whereas the corresponding values for POT and TPOT are 9.51\% and 6.20\%. Thus, even under more complex multiclass structures, POT provides more accurate estimates through geometric matching between the output and reference label distributions, and TPOT further reduces the effect of anomalous transport costs through thresholding to achieve the lowest BMAE among the compared methods.

\begin{table}[t]
	\renewcommand{\arraystretch}{1.1}
	\caption{Effect of the number of classes $K$ on estimation performance. Values are BMAE (\%); lower values indicate more accurate estimation. The best result is shown in {\textbf{bold}}, and the second-best result is \underline{underlined}.}
	\label{loss_compar}
	\centering
	\setlength{\tabcolsep}{1.6mm}
	\begin{tabular}{c|ccccc|cc}
		\bottomrule
		\cmidrule{1-8} 
		\multirow{2}{*}{$K$} 
		& \multicolumn{5}{c|}{{\textbf{Baseline}}} 
		& \multicolumn{2}{c}{{\textbf{Ours}}} \\
		& {AC} & {IM} & {DoC} & {ATC-MC} & {ATC-NE} & {POT} & {TPOT} \\ 
		\bottomrule
		\cmidrule{1-8}
		2 & 32.17 & 31.97 & 31.45 & 27.09 & 27.09 & \underline{03.76} & \textbf{02.17} \\
		3 & 35.81 & 34.46 & 32.50 & 29.48 & 28.79 & \underline{04.67} & \textbf{02.86} \\
		4 & 36.96 & 35.85 & 33.74 & 30.42 & 29.09 & \underline{07.94} & \textbf{04.77} \\
		5 & 38.28 & 36.03 & 35.11 & 31.82 & 31.41 & \underline{09.51} & \textbf{06.20} \\
		\bottomrule
		\cmidrule{1-8}
	\end{tabular}
\end{table}

\begin{figure*}[t]
	\centering
	\includegraphics[width=1.0\textwidth, height=0.44\textwidth]{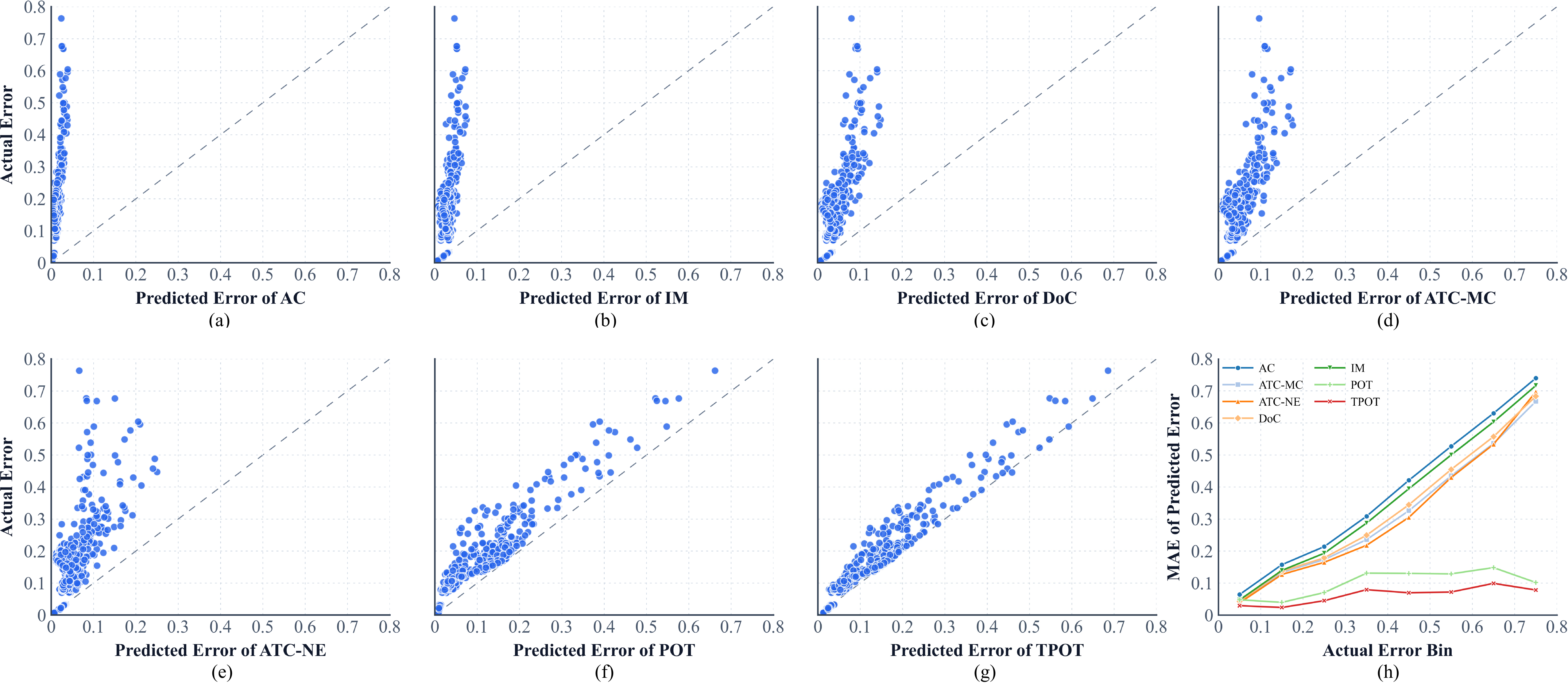}
	\vspace{-0.4cm}
	\caption{Error-rate estimates and interval-wise error curves of different label-free performance estimation methods in the five-class radar detection task. (a)--(g) show scatter plots of estimated versus true error rates for AC~\cite{ovadia2019can}, IM~\cite{liu2015classification}, DoC~\cite{guillory2021predicting}, ATC-MC~\cite{garg2022leveraging}, ATC-NE~\cite{garg2022leveraging}, POT, and TPOT, respectively. (h) shows the mean absolute error of each method over different true-error-rate intervals. In the multiclass setting, the baseline methods generally underestimate the true deployment error rate in medium- and high-error scenarios, whereas the POT and TPOT points are closer to the reference line and retain a clear advantage in BMAE.}
	\label{fig:ood_error_scatter_comp_5class}
\end{figure*}

\begin{figure*}[t]
	\centering
	\includegraphics[width=1.0\textwidth, height=0.23\textwidth]{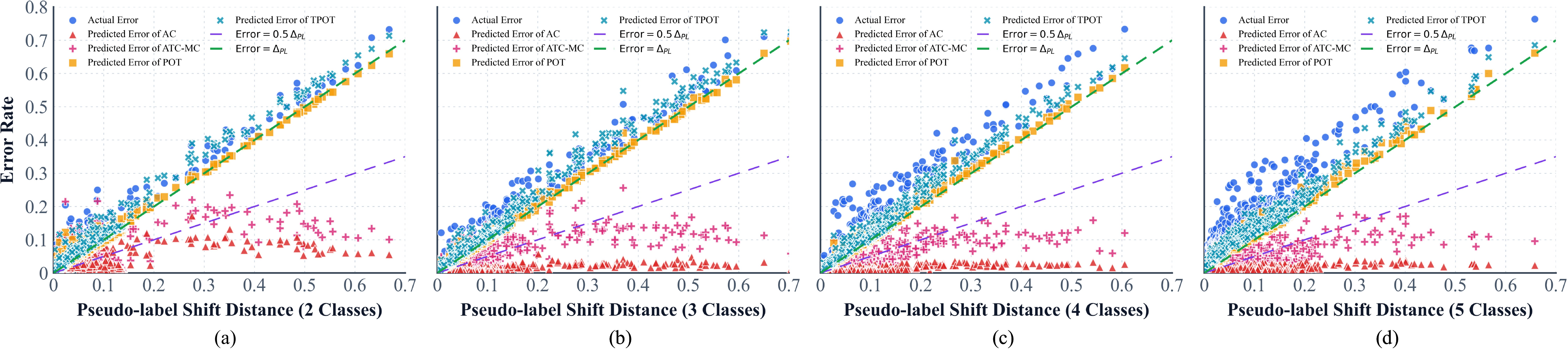}
	\vspace{-0.4cm}
	\caption{Relationship between error-rate estimates and pseudo-label shift for different numbers of classes. The horizontal axis is the pseudo-label shift distance $d_{\mathrm{PL}}$, and the vertical axis is the error rate. The green and purple dashed lines represent $\mathcal{E}=d_{\mathrm{PL}}$ and $\mathcal{E}=0.5d_{\mathrm{PL}}$, respectively. As the number of classes increases, many AC~\cite{ovadia2019can} and ATC-MC~\cite{garg2022leveraging} estimates fall below $0.5d_{\mathrm{PL}}$, indicating substantial underestimation. POT and TPOT remain largely above the theoretical lower bound and vary consistently with the true deployment error rate.}
	\label{fig:proposition3_exp}
\end{figure*}

Fig.~\ref{fig:ood_error_scatter_comp_5class} further presents the scatter plots and interval-wise error curves in the five-class setting. The AC, IM, DoC, ATC-MC, and ATC-NE points generally lie above and to the left of the $y=x$ reference line in the medium- and high-error regions, indicating that these methods continue to underestimate the true deployment error rate. Methods based on confidence-distribution differences or validation-calibrated thresholds may perform reasonably in low-error scenarios, but the correspondence between their score statistics and true error events can drift when the deployment distribution induces substantial class confusion. By contrast, the POT and TPOT points lie closer to the reference line, and their interval-wise MAE curves in Fig.~\ref{fig:ood_error_scatter_comp_5class}(h) are markedly below those of the baselines. The advantage of the proposed methods is therefore not restricted to binary classification but persists as the number of classes and the complexity of the confusion structure increase.

\subsubsection{Verification of the Propositions}
Finally, Fig.~\ref{fig:proposition3_exp} examines the theoretical property in Proposition~\ref{proposition3} from the perspective of pseudo-label shift. The four subfigures correspond to detection tasks with $K=2,3,4,5$. The true deployment error rate is generally greater than or equal to $d_{\mathrm{PL}}$, consistent with Proposition~\ref{proposition2}, which states that pseudo-label shift is a lower bound on the detection error rate. In many deployment scenarios, AC and ATC-MC estimates are far below the true deployment error rate, and as the number of classes increases, many points fall below $\mathcal{E}=0.5d_{\mathrm{PL}}$. Thus, estimators relying only on confidence or thresholded scores cannot consistently reflect the minimum risk of performance degradation implied by pseudo-label shift. By contrast, POT and TPOT estimates remain largely above $\mathcal{E}=0.5d_{\mathrm{PL}}$ and increase with $d_{\mathrm{PL}}$. This empirical behavior is consistent with the lower-bound property in Proposition~\ref{proposition3}, namely, $
{\mathcal{E}}_{\mathrm{POT}} \ge 0.5 W_\infty\left( {f}_{\#}P_{\mathrm{D}}\left(\hat{\mathbf{y}}\right), P_{\mathrm{D}}\left({\mathbf{y}}\right)\right).$
TPOT exhibits a trend similar to POT, indicating empirically that thresholding preserves POT's responsiveness to pseudo-label shift. Proposition~\ref{proposition3}, however, derives the theoretical lower bound only for POT; we therefore do not claim that TPOT has the same strict lower-bound guarantee. Overall, Fig.~\ref{fig:proposition3_exp} shows that the proposed methods are not only more accurate according to empirical error metrics but also consistent with the theoretically derived lower-bound relationship, providing stronger experimental support for their use in label-free online reliability estimation.

Taken together, POT and TPOT substantially outperform existing label-free performance estimation baselines in both binary and multiclass radar detection tasks. The results show that confidence-based methods tend to underestimate the true deployment error rate under distribution shift and pseudo-label shift, whereas optimal-transport-based distribution-level estimation more reliably captures the mismatch between the detector's output distribution and the training label structure. By thresholding transport costs, TPOT further improves robustness to anomalous outputs and is therefore more suitable as a performance-mismatch criterion for triggering model updates or self-evolutionary training in online radar detection systems.

\section{Conclusion}
\label{sec:conclusion}
This paper investigated label-free performance estimation for deep learning-based radar target detectors under distribution shift. From an optimal transport perspective, confidence-based estimators were shown to underestimate the true deployment error rate because they neglect the discrepancy between the pseudo-label and ground-truth label distributions. To address this limitation, we proposed probabilistic optimal transport (POT), which estimates the deployment error rate by measuring the Wasserstein distance between the detector's output probability distribution and the training label distribution. We further proposed thresholded probabilistic optimal transport (TPOT), which determines a transport-cost threshold from the validation set and reduces the influence of anomalous outputs through thresholding. Experiments on measured radar data showed that POT and TPOT estimate detection error rates more accurately across different distribution shifts and numbers of classes than existing label-free performance estimation methods.

A limitation of the proposed methods is that they rely primarily on marginal output and label distributions. Consequently, even when the training and deployment label priors are identical, error patterns that preserve the overall class proportions, such as systematic class permutations, may not be identifiable from unlabeled marginal distributions alone. Therefore, the current analysis does not establish universal identifiability of the deployment error rate. Future work will characterize additional conditions on the detector confusion structure and class-conditional output distributions, incorporate instance-level or conditional distribution information, and evaluate controlled scenarios in which the class priors remain unchanged while the detection error rate increases substantially.

\appendices 
\setcounter{proposition}{0} 
\setcounter{corollary}{0}
\setcounter{remark}{0}
{
	\section{PROOF OF PROPOSITION 1}\label{proof_1}
	
	\begin{proposition}
		Let $c_\infty(\mathbf{p},\mathbf{y})=\left\|\mathbf{p}-\mathbf{y}\right\|_\infty$ denote the transport cost between probability vector $\mathbf{p}\in\Delta^{K-1}$ output by $f$ and label vector $\mathbf{y}$, where $\left\|\cdot \right\|_\infty$ is the maximum norm, and let $W_{\infty}(\cdot,\cdot)$ denote the Wasserstein distance induced by this cost function. Then
		\begin{equation}
			{\mathcal{E}}_{\mathrm{AC}}=W_\infty\left( f_{\#}P_{\mathrm{D}}\left( \mathbf{p}\right), {f}_{\#}P_{\mathrm{D}}\left(\hat{\mathbf{y}}\right) \right).
		\end{equation}
	\end{proposition}
	
	\noindent{\bf{Proof:}} We first establish the following equality. Let $j^{\ast} = \mathrm{dec}\left(f\left(\mathbf{x}\right)\right)$, and let $\left[f\left(\mathbf{x}\right)\right]_{-j^{\ast}}$ denote the vector obtained by removing the $j^{\ast}$-th element from $f\left(\mathbf{x}\right)$. Then
	\begin{equation} 
		1 - \left[f\left(\mathbf{x}\right)\right]_{j^{\ast}} = \left\|\left[f\left(\mathbf{x}\right)\right]_{-j^{\ast}}\right\|_1 \ge \left\|\left[f\left(\mathbf{x}\right)\right]_{-j^{\ast}}\right\|_\infty .
	\end{equation}
	Therefore,
	\begin{equation}  
		\begin{aligned}
			\left\|\mathbf{y} \!-\! f\left(\mathbf{x}\right)\right\|_\infty &\!=\! \max\!\left\{\left\|\left[f\left(\mathbf{x}\right)\right]_{-j^{\ast}}\right\|_\infty, 1\! -\! \left[f\left(\mathbf{x}\right)\right]_{j^{\ast}}\!\right\} \\
			&= 1 - \left[f\left(\mathbf{x}\right)\right]_{j^{\ast}}.
		\end{aligned}
		\label{eq:proof1_1}
	\end{equation}
	
	Next, consider the OT plan between $f_{\#}P_{\mathrm{D}}\left( \mathbf{p}\right)$ and ${f}_{\#}P_{\mathrm{D}}\left(\hat{\mathbf{y}}\right)$. For any sample $\mathbf{x}^{(i)}$, the detector
	output probability vector $f(\mathbf{x}^{(i)})$ has a pseudo-label $\hat{\mathbf{y}}^{(i)}$ given by the one-hot vector corresponding to its largest probability component. Thus, among all possible label vectors, $\hat{\mathbf{y}}^{(i)}$ has the smallest $L_\infty$ distance to $f(\mathbf{x}^{(i)})$, i.e.,
	\begin{equation}  
		\begin{aligned}
			&\left\|f(\mathbf{x}^{(i)}) - \hat{\mathbf{y}}^{(i)}\right\|_\infty \le \left\|f(\mathbf{x}^{(i)}) - \tilde {\mathbf{y}}\right\|_\infty, \\
			&\qquad\quad \forall \tilde{\mathbf{y}} \in \{0, 1\}^K \cap \Delta^{K-1}. 
		\end{aligned}
	\end{equation} 
	In other words, for each output probability vector produced by the detector, its self-generated pseudo-label is the matching target with minimum transport cost. Moreover, ${f}_{\#}P_{\mathrm{D}}(\hat{\mathbf{y}})$ is precisely the empirical distribution formed by these pseudo-labels. Assigning each $f(\mathbf{x}^{(i)})$ to its corresponding $\hat{\mathbf{y}}^{(i)}$ therefore defines a feasible transport plan. Because this plan uses the minimum samplewise transport cost for every sample, its total transport cost is also minimal. Hence, when estimating the error rate of detector $f$ on $\mathcal{D}_{\mathrm{dep}}$,
	\begin{equation}
		W_{\!\infty}\!\left( f_{\#}\!P_{\mathrm{D}}\!\left( \mathbf{p}\right)\!, {f}_{\#}\!P_{\mathrm{D}}\!\left(\hat{\mathbf{y}}\right) \right) \!=\!\! \frac{1}{N_{\mathrm{dep}}}\!\! \sum_{i=1}^{N_{\mathrm{dep}}} \!\left\|f\left( \mathbf{x}^{(i)}\right) \!-\! \hat{\mathbf{y}}^{(i)}\right\|_\infty. 
		\label{eq:proof1_2}
	\end{equation}
	Combining~\eqref{eq:proof1_1} and~\eqref{eq:proof1_2} yields the following relationship between the AC-estimated error rate and the $W_\infty$ distance:
	\begin{equation}
		\begin{aligned}
			{\mathcal{E}}_{\mathrm{AC}} &= \frac{1}{N_{\mathrm{dep}}} \sum_{i=1}^{N_{\mathrm{dep}}} \left(1 - \max\left(  f\left( \mathbf{x}^{(i)}\right) \right)\right)  \\
			&= \frac{1}{N_{\mathrm{dep}}} \sum_{i=1}^{N_{\mathrm{dep}}} \left\| f\left( \mathbf{x}^{(i)}\right)  - \hat{\mathbf{y}}^{(i)}\right\|_\infty \\
			&= W_\infty\left( f_{\#}P_{\mathrm{D}}\left( \mathbf{p}\right), {f}_{\#}P_{\mathrm{D}}\left(\hat{\mathbf{y}}\right) \right).
		\end{aligned}
	\end{equation}
	
	This completes the proof. $\hfill\blacksquare$
	
	\section{PROOF OF COROLLARY 1}\label{proof_2}
	
	\begin{corollary}
		Let $P(\mathbf{y})$ be an arbitrary label distribution. Then
		\begin{equation}
			W_\infty\!\left( {f}_{\#}P_{\mathrm{D}}\!\left( \mathbf{p}\right)\!, P(\mathbf{y})\right)  \!\ge\! W_\infty\!\left( {f}_{\#}P_{\mathrm{D}}\!\left( \mathbf{p}\right)\!, {f}_{\#}P_{\mathrm{D}}\!\left( \hat{\mathbf{y}}\right) \right).\!
		\end{equation}
	\end{corollary}
	
	\noindent{\bf{Proof:}}
	From the main text,
	\begin{equation}  
		{\mathcal{E}}_{\mathrm{AC}} = 1 - \frac{1}{N_{\mathrm{dep}}} \sum_{i=1}^{N_{\mathrm{dep}}} \max\left( f(\mathbf{x}^{(i)})\right).
	\end{equation}
	Recall that, $\delta_0$ denotes the degenerate distribution with all probability mass concentrated at the zero vector, and $W_\infty(f_{\#}P_{\mathrm{D}}\left( \mathbf{p}\right), \delta_0)$ is the minimum cost of transporting $f_{\#}P_{\mathrm{D}}\left( \mathbf{p}\right)$ to $\delta_0$.
	Because $\delta_0$ has probability mass only at the zero vector, the unique matching target for every output probability vector produced by the detector $f\left( \mathbf{x}^{(i)}\right) $ is the zero vector. The transport cost of each $f\left( \mathbf{x}^{(i)}\right)$ is therefore $\left\|f(\mathbf{x}^{(i)}) - 0\right\|_\infty$. Hence, the estimated accuracy of detector $f$ on $\mathcal{D}_{\mathrm{dep}}$ is
	\begin{equation}  
		\begin{aligned}
			1 - {\mathcal{E}}_{\mathrm{AC}} &= \frac{1}{N_{\mathrm{dep}}} \sum_{i=1}^{N_{\mathrm{dep}}} \max\left( f(\mathbf{x}^{(i)})\right)  \\
			&= \frac{1}{N_{\mathrm{dep}}} \sum_{i=1}^{N_{\mathrm{dep}}} \left\|f(\mathbf{x}^{(i)}) - 0\right\|_\infty \\
			&= W_\infty(f_{\#}P_{\mathrm{D}}\left( \mathbf{p}\right), \delta_0) .
		\end{aligned}
		\label{eq:proof2_1}
	\end{equation}
	By the triangle inequality for the Wasserstein distance,
	\begin{equation}  
		\begin{aligned}
			W_\infty(P({\mathbf{y}}),\: f_{\#}P_{\mathrm{D}}\left( \mathbf{p}\right)) &+ W_\infty(f_{\#}P_{\mathrm{D}}\left( \mathbf{p}\right),\: \delta_0) \\
			&\ge W_\infty(P({\mathbf{y}}),\: \delta_0).
		\end{aligned}
	\end{equation}
	On the other hand, because $P(\mathbf{y})$ is defined over one-hot label vectors and every label vector $\mathbf{y}$ satisfies $|\mathbf{y}-0|_\infty=1$,
	\begin{equation}
		W_\infty(P(\mathbf{y}),\delta_0)=1 .
	\end{equation}
	Combining this result with~\eqref{eq:proof2_1} gives
	\begin{equation}  
		\begin{aligned}
			W_{\!\infty}(f_{\#}P_{\mathrm{D}}\left( \mathbf{p}\right), {f}_{\#}P_{\text{D}}(\hat{\mathbf{y}})) &= 1 - W_{\!\infty}(f_{\#}P_{\mathrm{D}}\left( \mathbf{p}\right), \delta_0) \\
			&\le W_{\!\infty}(f_{\#}P_{\mathrm{D}}\left( \mathbf{p}\right), P({\mathbf{y}})). 
		\end{aligned}
	\end{equation}
	
	This completes the proof. $\hfill\blacksquare$

	\section{PROOF OF PROPOSITION 2}\label{proof_3}
	\begin{proposition}
		Pseudo-label shift is a lower bound on the detector's true error rate:
		\begin{equation}
			W_\infty\left({f}_{\#}P_{\mathrm{D}}\left( \hat{\mathbf{y}}\right), P_\mathrm{D}(\mathbf{y})\right)
			\le \mathcal{E}_{\mathrm{D}} \le 1 .
		\end{equation}
	\end{proposition}
	
	\noindent{\bf{Proof:}} By definition, the true deployment error rate of the detector under the deployment distribution $P_{\mathrm{D}}(\mathbf{x},\mathbf{y})$ is
	\begin{equation}
		\mathcal{E}_{\mathrm{D}} = \mathbb{P}_{(\mathbf{x},\mathbf{y})\sim P_{\mathrm{D}}} \left( \hat{\mathbf{y}}\neq \mathbf{y} \right).
	\end{equation}
	Because $\hat{\mathbf{y}}$ and $\mathbf{y}$ are both one-hot vectors, if $\hat{\mathbf{y}}=\mathbf{y}$, then
	\begin{equation}
		\left\|\hat{\mathbf{y}}-\mathbf{y}\right\|_\infty=0.
	\end{equation}
	If $\hat{\mathbf{y}}\neq\mathbf{y}$, one class entry changes from 1 to 0 and another changes from 0 to 1, and hence
	\begin{equation}
		\left\|\hat{\mathbf{y}}-\mathbf{y}\right\|_\infty=1.
	\end{equation}
	Therefore,
	\begin{equation}
		\left\|\hat{\mathbf{y}}-\mathbf{y}\right\|_\infty
		=
		\mathbb{I}\left(\hat{\mathbf{y}}\neq\mathbf{y}\right).
	\end{equation}
	Taking the expectation over the deployment distribution gives
	\begin{equation}
		\mathcal{E}_{\mathrm{D}}
		=
		\mathbb{E}_{(\mathbf{x},\mathbf{y})\sim P_{\mathrm{D}}}
		\left[
		\left\|\hat{\mathbf{y}}-\mathbf{y}\right\|_\infty
		\right].
	\end{equation}
	The samplewise pairing induced by the joint distribution of $(\hat{\mathbf{y}},\mathbf{y})$ is a feasible transport plan between ${f}_{\#}P_{\mathrm{D}}(\hat{\mathbf{y}})$ and $P_{\mathrm{D}}(\mathbf{y})$, because its two marginals are the pseudo-label and ground-truth label distributions. The Wasserstein distance minimizes the expected transport cost over all feasible transport plans and therefore cannot exceed the cost under any particular feasible plan. It follows that
	\begin{equation}
		W_{\!\infty}\!\left({f}_{\#}P_{\mathrm{D}}\!\left( \hat{\mathbf{y}}\right), P_\mathrm{D}(\mathbf{y})\right)
		\!\le\! \mathbb{E}_{(\mathbf{x},\mathbf{y})\sim P_{\mathrm{D}}} \left[ \left\|\hat{\mathbf{y}}\!-\!\mathbf{y}\right\|_{\infty} \right] \!=\! \mathcal{E}_{\mathrm{D}}.
	\end{equation}
	Moreover, because $\mathcal{E}_{\mathrm{D}}$ is an error probability, it clearly satisfies $0\le\mathcal{E}_{\mathrm{D}}\le1$. Therefore,
	\begin{equation}
		W_\infty\left({f}_{\#}P_{\mathrm{D}}\left( \hat{\mathbf{y}}\right), P_\mathrm{D}(\mathbf{y})\right)
		\le
		\mathcal{E}_{\mathrm{D}}
		\le
		1 .
	\end{equation}
	
	This completes the proof. $\hfill\blacksquare$
	
	\section{PROOF OF PROPOSITION 3}\label{proof_4}
	For any $\mathbf{p}^{(i)} \in \Delta^{K-1}$, define $\mathcal{C}(\mathbf{p}^{(i)}) = \{\tilde{\mathbf{p}} \in \Delta^{K-1} \mid \mathrm{dec}\left( \tilde{\mathbf{p}}\right) = \mathrm{dec}\left( \mathbf{p}^{(i)}\right)\}$ to be the set of probability vectors having the same pseudo-label as $\mathbf{p}^{(i)}$. The set $\mathcal{C}(\mathbf{p}^{(i)})$ can also be interpreted as the equivalence class determined by that pseudo-label. Further, let $\mathcal{P}(\Delta^{K-1})$ be the set of all probability-vector distributions defined on the probability simplex $\Delta^{K-1}$. For a given pseudo-label distribution ${f}_{\#}P_{\mathrm{D}}\left( \hat{\mathbf{y}}\right)$, define
	\begin{equation}
		\begin{aligned}
			\mathcal{P}_{\mathbf{p}}[{f}_{\#}P_{\mathrm{D}}\left( \hat{\mathbf{y}}\right)] =& \{P(\mathbf{p}) \in \mathcal{P}(\Delta^{K-1}) \mid \\
			&P\left( \mathrm{dec} \left( \mathbf{p}\right)  = \mathrm{dec}\left(  \hat{\mathbf{y}}\right)\right)  = {f}_{\#}P_{\mathrm{D}}\left( \hat{\mathbf{y}}\right) \}.
		\end{aligned}
	\end{equation}
	This set contains all probability-vector distributions that induce the same pseudo-label distribution ${f}_{\#}P_{\mathrm{D}}\left( \hat{\mathbf{y}}\right)$. In Fig.~\ref{fig:AC_Wasserstein}, $\mathcal{P}_{\mathbf{p}}[{f}_{\#}P_{\mathrm{D}}\left( \hat{\mathbf{y}}\right)]$ can be illustrated by the line segment between $\delta_0$ and ${f}_{\#}P_{\mathrm{D}}\left( \hat{\mathbf{y}}\right)$; every probability-vector distribution $P(\mathbf{p})$ on this segment shares the same pseudo-label distribution ${f}_{\#}P_{\mathrm{D}}\left( \hat{\mathbf{y}}\right)$.
	
	We first introduce a lemma to prove Proposition~3.
	\begin{lemma}
		Let ${\mathbf{y}}, {\mathbf{y}}' \in \{0, 1\}^K \cap \Delta^{K-1}$ be two one-hot label vectors. Then
		\begin{equation} 
			\inf_{\mathbf{p} \in \mathcal{C}({\mathbf{y}})} \left\|\mathbf{p} - {\mathbf{y}}'\right\|_\infty = 0.5 \times \mathbb{I} \left( {\mathbf{y}} \neq {\mathbf{y}}'\right) . 
			\label{eq:lemma1} 
		\end{equation}
	\end{lemma}
	\begin{remark}
		This lemma states that, when only the decision class of a probability vector is considered, the minimum $\ell_\infty$ distance from a probability-vector equivalence class to a one-hot label point depends only on whether their classes agree.
	\end{remark}
	
	\noindent{\bf{Proof:}} If ${\mathbf{y}}={\mathbf{y}}'$, then ${\mathbf{y}}\in \mathcal{C}({\mathbf{y}})$, and therefore
	\begin{equation}  
		\inf_{\mathbf{p} \in \mathcal{C}({\mathbf{y}})} \left\|\mathbf{p} - {\mathbf{y}}'\right\|_\infty \le \left\|{\mathbf{y}} - {\mathbf{y}}'\right\|_\infty = 0.
	\end{equation}
	Because $\left\|\cdot \right\|_\infty$ is nonnegative,
	\begin{equation}
		\inf_{\mathbf{p} \in \mathcal{C}({\mathbf{y}})} \left\|\mathbf{p} - {\mathbf{y}}'\right\|_\infty = 0.
	\end{equation}
	
	Now consider ${\mathbf{y}}\neq{\mathbf{y}}'$. Let $a$ and $b$ be the class indices associated with ${\mathbf{y}}$ and ${\mathbf{y}}'$, respectively, where $a\neq b$. We first establish an upper bound on $\inf_{\mathbf{p} \in \mathcal{C}({\mathbf{y}})} \left\|\mathbf{p} - {\mathbf{y}}'\right\|_\infty$. For any $\lambda\in(0,0.5]$, construct
	\begin{equation}
		\mathbf{p}_{\lambda} = (0.5+\lambda){\mathbf{y}} + (0.5-\lambda){\mathbf{y}}' .
	\end{equation}
	Then $\mathbf{p}_{\lambda}\in\Delta^{K-1}$, assigning probabilit $0.5+\lambda$ to class $a$ and $0.5-\lambda$ for class $b$. Therefore,
	\begin{equation}
		\mathrm{dec}(\mathbf{p}_{\lambda})=a=\mathrm{dec}({\mathbf{y}}).
	\end{equation}
	Thus, $\mathbf{p}_{\lambda}\in\mathcal{C}({\mathbf{y}})$. Moreover,
	\begin{equation}
		\left\|\mathbf{p}_{\lambda}-{\mathbf{y}}'\right\|_\infty = 0.5+\lambda.
	\end{equation}
	Letting $\lambda\rightarrow 0^+$ gives
	\begin{equation}
		\inf_{\mathbf{p} \in \mathcal{C}({\mathbf{y}})} \left\|\mathbf{p}-{\mathbf{y}}'\right\|_\infty \le 0.5 .
	\end{equation}
	
	We next establish a lower bound on $\inf_{\mathbf{p} \in \mathcal{C}({\mathbf{y}})} \left\|\mathbf{p} - {\mathbf{y}}'\right\|_\infty$. Suppose that some $\mathbf{p}\in\Delta^{K-1}$ satisfies
	\begin{equation}
		\left\|\mathbf{p}-{\mathbf{y}}'\right\|_\infty <0.5,
	\end{equation}
	Because the class associated with ${\mathbf{y}}'$ is $b$, it follows that
	\begin{equation}
		\mathrm{dec}(\mathbf{p})=b=\mathrm{dec}({\mathbf{y}}')\neq \mathrm{dec}({\mathbf{y}}).
	\end{equation}
	Thus, every $\mathbf{p}$ satisfying $\left\|\mathbf{p}-{\mathbf{y}}'\right\|_\infty <0.5$ lies outside $\mathcal{C}({\mathbf{y}})$. Therefore,
	\begin{equation}
		\inf_{\mathbf{p} \in \mathcal{C}({\mathbf{y}})} \left\|\mathbf{p}-{\mathbf{y}}'\right\|_\infty \ge 0.5 .
	\end{equation}
	
	Combining the upper and lower bounds, when ${\mathbf{y}}\neq{\mathbf{y}}'$,
	\begin{equation}
		\inf_{\mathbf{p} \in \mathcal{C}({\mathbf{y}})} \left\|\mathbf{p}-{\mathbf{y}}'\right\|_\infty =0.5 .
	\end{equation}
	This completes the proof of the lemma. $\hfill\blacksquare$
	
	\begin{proposition}\label{ap_proposition3}
		Under the assumption $P_{\mathrm{T}}\left({\mathbf{y}}\right) = P_{\mathrm{D}}\left({\mathbf{y}}\right)$, the following inequality always holds:
		\begin{equation} 
			{\mathcal{E}}_{\mathrm{POT}} \ge 0.5 W_\infty\left( {f}_{\#}P_{\mathrm{D}}\left(\hat{\mathbf{y}}\right), P_{\mathrm{D}}\left({\mathbf{y}}\right)\right). 
		\end{equation}
	\end{proposition}
	
	\noindent{\bf{Proof:}} Under $P_{\mathrm{T}}\left({\mathbf{y}}\right) = P_{\mathrm{D}}\left({\mathbf{y}}\right)$, the POT estimator can be written as
	\begin{equation} 
		\begin{aligned} 
			&{\mathcal{E}}_{\text{POT}} = W_\infty(f_\# P_{\mathrm{D}}(\mathbf{p}), P_{\mathrm{D}}({\mathbf{y}})) \\
			&= \inf_{\pi(\mathbf{p}, {\mathbf{y}}) \in \Pi(f_\# P_{\mathrm{D}}(\mathbf{p}), P_{\mathrm{D}}({\mathbf{y}}))} \int \left\|\mathbf{p} - {\mathbf{y}}\right\|_\infty d\pi(\mathbf{p}, {\mathbf{y}}) \\
			&\ge \inf_{\pi(\mathbf{p}, {\mathbf{y}}) \in \Pi(f_\# P_{\mathrm{D}}(\mathbf{p}), P_{\mathrm{D}}({\mathbf{y}}))} \int \inf_{\tilde{\mathbf{p}} \in \mathcal{C}(\mathbf{p})} \left\|\tilde{\mathbf{p}} - {\mathbf{y}}\right\|_\infty d\pi(\mathbf{p}, {\mathbf{y}}) .
			\label{eq:prop3_ineq}
		\end{aligned}
	\end{equation}
	For a probability vector $\mathbf{p}$ and its corresponding pseudo-label $\hat{\mathbf{y}}$, $\mathcal{C}(\mathbf{p}) = \mathcal{C}(\hat{\mathbf{y}})$. Therefore, $\inf_{\tilde{\mathbf{p}} \in \mathcal{C}(\hat{\mathbf{y}})} \left\|\tilde{\mathbf{p}} - {\mathbf{y}}\right\|_\infty$ depends only on the pseudo-label $\hat{\mathbf{y}}$ and the ground-truth label ${\mathbf{y}}$, not on the specific probability values of $\mathbf{p}$ within the decision class. The integral over $\mathbf{p}$ and ${\mathbf{y}}$ can thus be rewritten as an integral over $\hat{\mathbf{y}}$ and ${\mathbf{y}}$, yielding
	\begin{equation} 
		{\mathcal{E}}_{\text{POT}} \!\ge\!\!\! \inf_{\pi(\hat{\mathbf{y}}, {\mathbf{y}}) \in \Pi({f}_{\#}P_{\mathrm{D}}\left( \hat{\mathbf{y}}\right), P_{\mathrm{D}}({\mathbf{y}}))} \!\int\!\! \inf_{\tilde{\mathbf{p}} \in \mathcal{C}(\hat{\mathbf{y}})} \!\!\left\|\tilde{\mathbf{p}} \!-\! {\mathbf{y}}\right\|_\infty d\pi(\hat{\mathbf{y}}, {\mathbf{y}}).
		\label{eq:proposition3}
	\end{equation}
	Substituting Lemma~1 into~\eqref{eq:proposition3} gives
	\begin{equation} 
		\begin{aligned}
			{\mathcal{E}}_{\text{POT}} \ge& \inf_{\pi(\hat{\mathbf{y}}, {\mathbf{y}}) \in \Pi({f}_{\#}P_{\mathrm{D}}\left( \hat{\mathbf{y}}\right), P_{\mathrm{D}}({\mathbf{y}}))} \int \frac{1}{2} \mathbb{I}\left( \hat{\mathbf{y}} \neq {\mathbf{y}}\right)  d\pi(\hat{\mathbf{y}}, {\mathbf{y}}) \\
			=& \frac{1}{2} \inf_{\pi(\hat{\mathbf{y}}, {\mathbf{y}}) \in \Pi({f}_{\#}P_{\mathrm{D}}\left( \hat{\mathbf{y}}\right), P_{\mathrm{D}}({\mathbf{y}}))} \int \|\hat{\mathbf{y}} - {\mathbf{y}}\|_\infty d\pi(\hat{\mathbf{y}}, {\mathbf{y}}) \\
			=& 0.5 W_\infty({f}_{\#}P_{\mathrm{D}}\left( \hat{\mathbf{y}}\right), P_{\mathrm{D}}({\mathbf{y}})). 
		\end{aligned}
	\end{equation}
	
	This completes the proof. $\hfill\blacksquare$

	\section{PROOF OF THE TIGHTNESS OF THE LOWER BOUND IN PROPOSITION 3}\label{proof_5}
	Although inequality~\eqref{eq:prop3_ineq} may appear loose, the lower bound is in fact tight without further assumptions on the calibration of classifier $f$. We establish this tightness using a lemma and a corollary.
	\begin{lemma}\label{lemma2}
		Pseudo-label shift is related to the total variation distance between the target label distribution and the pseudo-label distribution as follows:
		\begin{equation} 
			W_\infty({f}_{\#}P_{\mathrm{D}}\left( \hat{\mathbf{y}}\right), P_{\mathrm{D}}({\mathbf{y}})) = \|{f}_{\#}P_{\mathrm{D}}\left( \hat{\mathbf{y}}\right) - P_{\mathrm{D}}({\mathbf{y}})\|_{\mathrm{TV}}.
		\end{equation}
	\end{lemma}
	\begin{remark}
		This result is established in~\cite{villani2009optimal}. The equivalence is meaningful specifically when measuring the $W_\infty$ distance between two label distributions.
	\end{remark}
	
	\begin{corollary}
		\begin{equation} 
			W_\infty({f}_{\#}P_{\mathrm{D}}\left( \hat{\mathbf{y}}\right)\!, 	P_{\mathrm{D}}({\mathbf{y}})) \!=\! 1 \!-\! \sum\! \min\{{f}_{\#}P_{\mathrm{D}}\left( \hat{\mathbf{y}}\right)\!, P_{\mathrm{D}}({\mathbf{y}})\} .
		\end{equation}
	\end{corollary}
	
	\noindent{\bf{Proof:}}
	\begin{equation} 
		\begin{aligned}
			&W_\infty({f}_{\#}P_{\mathrm{D}}\left( \hat{\mathbf{y}}\right), P_{\mathrm{D}}({\mathbf{y}}))\\ 
			&= \|{f}_{\#}P_{\mathrm{D}}\left( \hat{\mathbf{y}}\right) \!-\! P_{\mathrm{D}}({\mathbf{y}})\|_{\mathrm{TV}} 
			= \frac{1}{2} \sum \left| {f}_{\#}P_{\mathrm{D}}\left( \hat{\mathbf{y}}\right) \!-\! P_{\mathrm{D}}({\mathbf{y}})\right|  \\
			& = \!\frac{1}{2}\! \sum\! \max\{{f}_{\#}P_{\mathrm{D}}\left( \hat{\mathbf{y}}\right), P_{\mathrm{D}}({\mathbf{y}})\} \!-\! \min\{{f}_{\#}P_{\mathrm{D}}\left( \hat{\mathbf{y}}\right), P_{\mathrm{D}}({\mathbf{y}})\} \\
			& = \!\frac{1}{2}\! \sum\! \max\{{f}_{\#}P_{\mathrm{D}}\left( \hat{\mathbf{y}}\right), P_{\mathrm{D}}({\mathbf{y}})\} \!+\! \min\{{f}_{\#}P_{\mathrm{D}}\left( \hat{\mathbf{y}}\right), P_{\mathrm{D}}({\mathbf{y}})\} \\
			& \quad-\! \sum \min\{{f}_{\#}P_{\mathrm{D}}\left( \hat{\mathbf{y}}\right), P_{\mathrm{D}}({\mathbf{y}})\} \\
			& = \!\frac{1}{2}\! \sum\! {f}_{\#}P_{\mathrm{D}}\left( \hat{\mathbf{y}}\right) \!+\! P_{\mathrm{D}}({\mathbf{y}}) \!-\! \sum\! \min\{{f}_{\#}P_{\mathrm{D}}\left( \hat{\mathbf{y}}\right), P_{\mathrm{D}}({\mathbf{y}})\} \\
			& = 1 \!-\! \sum \min\{{f}_{\#}P_{\mathrm{D}}\left( \hat{\mathbf{y}}\right), P_{\mathrm{D}}({\mathbf{y}})\}. 
		\end{aligned}
	\end{equation}
	This completes the proof of the corollary.
	
	We next prove that the lower bound is tight by constructing a sequence of probability-vector distributions whose limiting distance to $P_{\mathrm{D}}({\mathbf{y}})$ is exactly $0.5 W_\infty({f}_{\#}P_{\mathrm{D}}\left( \hat{\mathbf{y}}\right), P_{\mathrm{D}}({\mathbf{y}}))$.
	
	\begin{lemma}\label{lemma3}
		The lower bound in Proposition~\ref{ap_proposition3} is tight:
		\begin{equation}  
			\inf_{\!\! P\!(\mathbf{p}) \in \mathcal{P}_{\mathbf{p}}[{f}_{\#}\!P_{\mathrm{D}}\left( \hat{\mathbf{y}}\right)]} \!\!\!\!\!\!\!W_{\!\infty}(P(\mathbf{p}),\! P_{\mathrm{D}}({\mathbf{y}})) \!=\!\! \frac{1}{2} W_{\!\infty}({f}_{\#}P_{\mathrm{D}}\!\left( \hat{\mathbf{y}}\right)\!,\! P_{\mathrm{D}}({\mathbf{y}})). 
		\end{equation}
	\end{lemma}
	
	\noindent{\bf{Proof:}} We first construct the family of distributions $\{P_\lambda(\mathbf{p}) \mid \lambda \in \left( 0, 0.5\right] \}$, where $P_\lambda(\mathbf{p})$ is defined as the following mixture distribution:
	\begin{equation} 
		P_\lambda(\mathbf{p}) = \omega P_{\mathrm{ov}}(\hat{\mathbf{y}}) + (1 - \omega) P_{\mathrm{res}}(\mathbf{p}), 
	\end{equation}
	where $P_{\mathrm{ov}}({\mathbf{y}}) = \omega^{-1} \min\{{f}_{\#}P_{\mathrm{D}}\left( \hat{\mathbf{y}}\right), P_{\mathrm{D}}({\mathbf{y}})\}$, $\omega = \sum \min\{{f}_{\#}P_{\mathrm{D}}\left( \hat{\mathbf{y}}\right), P_{\mathrm{D}}({\mathbf{y}})\}$, and $P_{\mathrm{res}}(\mathbf{p})$ is supported on $\Delta^{K-1} \cap \{0.5 + \lambda, 0.5 - \lambda, 0\}^K$, i.e., one element of $\mathbf{p}$ is $0.5 + \lambda$, another is $0.5 - \lambda$, and all remaining elements are zero. In addition, $P_{\mathrm{res}}(\left[ \mathbf{p}\right]_i = 0.5 + \lambda) = {f}_{\#}P_{\mathrm{D}}(\hat{\mathbf{y}}_i = 1)$ and $P_{\mathrm{res}}(\left[ \mathbf{p}\right]_i = 0.5 - \lambda) = P_{\mathrm{D}}(\hat{\mathbf{y}}_i = 1)$. It is straightforward to verify that $P_\lambda(\mathbf{p}) \in \mathcal{P}_{\mathbf{p}}[{f}_{\#}P_{\mathrm{D}}\left( \hat{\mathbf{y}}\right)]$.
	
	Next, we construct an explicit transport plan $\pi(\mathbf{p}, {\mathbf{y}}) \in \Pi(P_\lambda(\mathbf{p}), P_{\mathrm{D}}({\mathbf{y}}))$ through the factorization $\pi(\mathbf{p}, {\mathbf{y}}) = P_\lambda(\mathbf{p})\pi({\mathbf{y}}|\mathbf{p})$, where
	\begin{equation} 
		\pi({\mathbf{y}}|\mathbf{p}) = \begin{cases} 1 & \text{if } \mathbf{p} = {\mathbf{y}} \text{ or } \langle \mathbf{p}, {\mathbf{y}} \rangle = 0.5 - \lambda \\ 0 & \text{otherwise} \end{cases} .
	\end{equation}
	The cost of this transport plan is therefore
	\begin{equation} 
		\begin{aligned}
			\int \|\mathbf{p} - {\mathbf{y}}\|_\infty &d\pi(\mathbf{p}, {\mathbf{y}}) = (0.5 + \lambda)(1 - \omega) \\
			&= (0.5 + \lambda) W_\infty({f}_{\#}P_{\mathrm{D}}\left( \hat{\mathbf{y}}\right), P_{\mathrm{D}}({\mathbf{y}})) ,
		\end{aligned}
	\end{equation}
	where the last equality follows from Lemma~\ref{lemma2}. Taking the infimum gives
	\begin{equation} 
		\inf_{\lambda \in (0, 0.5]} \!\int\!\! \|\mathbf{p} \!-\! {\mathbf{y}}\|_\infty d\pi(\mathbf{p}, {\mathbf{y}}) \!=\! \frac{1}{2} W_\infty({f}_{\#}P_{\mathrm{D}}\left( \hat{\mathbf{y}}\right), P_{\mathrm{D}}({\mathbf{y}})).
	\end{equation}
	Combining the preceding results yields
	\begin{align}
			&0.5 W_\infty({f}_{\#}P_{\mathrm{D}}\left( \hat{\mathbf{y}}\right), P_{\mathrm{D}}({\mathbf{y}}))\nonumber\\
			&\qquad= \inf_{\lambda \in (0, 0.5]} \int \|\mathbf{p} - {\mathbf{y}}\|_\infty d\pi(\mathbf{p}, {\mathbf{y}}) \nonumber\\
			&\qquad \ge \inf_{\lambda \in (0, 0.5]} W_\infty(P_\lambda(\mathbf{p}), P_{\mathrm{D}}({\mathbf{y}})) \label{eq_ineq1}\\
			&\qquad \ge \inf_{P(\mathbf{p}) \in \mathcal{P}_{\mathbf{p}}[{f}_{\#}P_{\mathrm{D}}\left( \hat{\mathbf{y}}\right)]} 	W_\infty(P(\mathbf{p}), P_{\mathrm{D}}({\mathbf{y}})) \label{eq_ineq2}\\
			&\qquad \ge 0.5 W_\infty({f}_{\#}P_{\mathrm{D}}\left( \hat{\mathbf{y}}\right), 	P_{\mathrm{D}}({\mathbf{y}})) \label{eq_ineq3}.
	\end{align}
	Inequality \eqref{eq_ineq1} follows because, for each $\lambda\in(0,0.5]$, the explicitly constructed transport plan
	$\pi(\mathbf{p},\mathbf{y})$ is a feasible coupling in 	$\Pi(P_\lambda(\mathbf{p}),P_{\mathrm{D}}(\mathbf{y}))$.
	Therefore, its transport cost cannot be smaller than the OT cost defining $W_\infty(P_\lambda(\mathbf{p}),P_{\mathrm{D}}(\mathbf{y}))$. Taking the infimum over $\lambda$ preserves this inequality. Inequality \eqref{eq_ineq2} follows from
	\begin{equation}
		\left\lbrace P_\lambda(\mathbf{p})\mid\lambda\in(0,0.5]\right\rbrace \subseteq \mathcal{P}_{\mathbf{p}}[{f}_{\#}P_{\mathrm{D}}(\hat{\mathbf{y}})].
	\end{equation}
	Specifically, taking the infimum over the entire admissible set cannot yield a value larger than taking the infimum over one of its subsets. Finally, inequality~\eqref{eq_ineq3} is an application of the lower bound
	established in Proposition~\ref{ap_proposition3}, which holds for every $P(\mathbf{p})\in \mathcal{P}_{\mathbf p}[{f}_{\#}P_{\mathrm{D}}(\hat{\mathbf{y}})]$.
	
	Because the first and last terms in the above inequality chain are identical,
	all intermediate inequalities must hold with equality. Therefore,
	\begin{equation}
		\inf_{P(\mathbf{p}) \in \mathcal{P}_{\mathbf{p}}[{f}_{\!\#}\!P_{\mathrm{D}}\left( \hat{\mathbf{y}}\right)]} \!\!\!\!\!\!W_{\!\infty}\!(P(\mathbf{p})\!, \!P_{\mathrm{D}}({\mathbf{y}})) \!=\! \frac{1}{2} W_{\!\infty}\!({f}_{\!\#}\!P_{\mathrm{D}}\!\left( \hat{\mathbf{y}}\right)\!, \!P_{\mathrm{D}}({\mathbf{y}})),
	\end{equation} 
	which completes the proof of Lemma~\ref{lemma3}. $\hfill\blacksquare$
}

\bibliographystyle{IEEEtran}
\bibliography{IEEEabrv,myref}

\begin{thebibliography}{10}
\providecommand{\url}[1]{#1}
\csname url@samestyle\endcsname
\providecommand{\newblock}{\relax}
\providecommand{\bibinfo}[2]{#2}
\providecommand{\BIBentrySTDinterwordspacing}{\spaceskip=0pt\relax}
\providecommand{\BIBentryALTinterwordstretchfactor}{4}
\providecommand{\BIBentryALTinterwordspacing}{\spaceskip=\fontdimen2\font plus
\BIBentryALTinterwordstretchfactor\fontdimen3\font minus
  \fontdimen4\font\relax}
\providecommand{\BIBforeignlanguage}[2]{{%
\expandafter\ifx\csname l@#1\endcsname\relax
\typeout{** WARNING: IEEEtran.bst: No hyphenation pattern has been}%
\typeout{** loaded for the language `#1'. Using the pattern for}%
\typeout{** the default language instead.}%
\else
\language=\csname l@#1\endcsname
\fi
#2}}
\providecommand{\BIBdecl}{\relax}
\BIBdecl

\bibitem{10535463}
A.~Zaimbashi, M.~S. Greco, and F.~Gini, ``Integrated mimo passive radar target
  detection,'' \emph{IEEE Trans. Signal Process.}, vol.~72, pp. 2677--2691,
  2024.

\bibitem{10675372}
X.~Wang, Y.~Wang, C.~Zang, X.~Chen, H.~Ren, and G.~Cui, ``Intelligent maritime
  radar target detection with partial annotation via progressive learning,''
  \emph{IEEE Sens. J.}, vol.~24, no.~21, pp. 34\,987--34\,998, 2024.

\bibitem{11543476}
X.~Du, D.~Ma, X.~Chen, G.~Cui, and J.~Zheng, ``Sea clutter suppression for
  maritime radar measurements with mpdnet,'' \emph{IEEE Trans. Aerosp.
  Electron. Syst.}, vol.~62, pp. 12\,070--12\,084, 2026.

\bibitem{10371148}
Y.~Wang, X.~Wang, C.~Zang, W.~Zhao, G.~Cui, and S.~Guo, ``Multi-polarization
  features fusion detection of marine small targets based on {LSTM},'' in
  \emph{2023 IEEE Int. Radar Conf. (RADAR)}, 2023, pp. 1--5.

\bibitem{10848402}
C.~Zang, Y.~Wang, X.~Wang, C.~Xu, and G.~Cui, ``Sea clutter suppression via
  cuttable encoder-decoder-augmentation network,'' \emph{J. Syst. Eng.
  Electron.}, vol.~35, no.~6, pp. 1428--1440, 2024.

\bibitem{10929070}
M.~Tang, A.~Aubry, A.~De~Maio, and Y.~Rong, ``Invariance theory for radar
  detection in disturbance with kronecker product covariance structure—part
  i: Gaussian environment,'' \emph{IEEE Trans. Signal Process.}, vol.~73, pp.
  1577--1593, 2025.

\bibitem{10879091}
Q.~Lv, L.~Liu, Y.~Wu, M.~Peng, and Y.~Quan, ``A framework for compound
  interrupted sampling repeater jamming parameter measurement,'' \emph{IEEE
  Trans. Instrum. Meas.}, vol.~74, pp. 1--4, 2025.

\bibitem{9807335}
S.~Chen, C.~Feng, Y.~Huang, X.~Chen, and F.~Li, ``Small target detection in
  x-band sea clutter using the visibility graph,'' \emph{IEEE Trans. Geosci.
  Remote Sens.}, vol.~60, pp. 1--11, 2022.

\bibitem{10742946}
T.-Y. Duan, P.-L. Shui, J.-M. Wang, and S.-W. Xu, ``Fusion-based adaptive
  coherent detection of small targets in dual-polarimetric correlated sea
  clutter,'' \emph{IEEE Trans. Aerosp. Electron. Syst.}, vol.~61, no.~2, pp.
  3868--3881, 2025.

\bibitem{11193878}
Z.~Xu, W.~Liu, X.~Chen, H.~Dai, J.~Liu, and H.~Chen, ``Distributed target
  detection in compound-gaussian clutter under steering vector uncertainty,''
  \emph{IEEE Trans. Aerosp. Electron. Syst.}, vol.~62, pp. 26--40, 2026.

\bibitem{11267251}
J.~Xue, Z.~Fan, S.~Xu, and M.~Pan, ``Persymmetric adaptive detection for radar
  targets in compound gaussian clutter with nakagami texture,'' \emph{IEEE
  Trans. Aerosp. Electron. Syst.}, vol.~62, pp. 2292--2303, 2026.

\bibitem{9145700}
J.~Liu, D.~Massaro, D.~Orlando, and A.~Farina, ``Radar adaptive detection
  architectures for heterogeneous environments,'' \emph{IEEE Trans. Signal
  Process.}, vol.~68, pp. 4307--4319, 2020.

\bibitem{10415070}
N.~Su, X.~Chen, J.~Guan, Y.~Huang, X.~Wang, and Y.~Xue, ``Radar maritime target
  detection via spatial–temporal feature attention graph convolutional
  network,'' \emph{IEEE Trans. Geosci. Remote Sens.}, vol.~62, pp. 1--15, 2024.

\bibitem{10371351}
Y.~Wang, C.~Zang, B.~Yu, W.~Zhao, X.~Wang, C.~Xu, and G.~Cui, ``{WTE-CGAN}
  based signal enhancement for weak target detection,'' \emph{IEEE Geosci.
  Remote Sens. Lett.}, vol.~21, pp. 1--5, 2024.

\bibitem{11511344}
J.~Xue, Z.~Feng, S.~Xu, and M.~Pan, ``Mvlsn-segnet: Multifeature view-based
  land–sea–noise segmentation network for maritime surveillance radars,''
  \emph{IEEE Trans. Radar Syst.}, vol.~4, pp. 943--955, 2026.

\bibitem{10530230}
Q.~Lv, H.~Fan, J.~Liu, Y.~Zhao, M.~Xing, and Y.~Quan, ``Multilabel deep
  learning-based lightweight radar compound jamming recognition method,''
  \emph{IEEE Trans. Instrum. Meas.}, vol.~73, pp. 1--15, 2024.

\bibitem{11425796}
J.~Xue, X.~Zhao, S.~Xu, X.~Bai, and M.~Pan, ``Fptf-yolo: Fully polarimetric
  time–frequency yolo for maritime radar weak target detection and
  localization,'' \emph{IEEE Trans. Aerosp. Electron. Syst.}, vol.~62, pp.
  7651--7664, 2026.

\bibitem{11125835}
X.~Chen, S.~Xu, S.~Hu, and X.~Ma, ``Uifd: A unified interactive network for
  image fusion and traffic object detection under low-light conditions,''
  \emph{IEEE Trans. Intell. Transp. Syst.}, vol.~26, no.~11, pp.
  19\,182--19\,196, 2025.

\bibitem{11328771}
J.-Y. Li, B.-W. Fang, P.-L. Shui, and S.-W. Xu, ``Open-set classification of
  maritime radar small targets using openmax layer and supervised variational
  autoencoder,'' \emph{IEEE Trans. Aerosp. Electron. Syst.}, vol.~62, pp.
  3991--4006, 2026.

\bibitem{9622244}
Q.~Lv, Y.~Quan, W.~Feng, M.~Sha, S.~Dong, and M.~Xing, ``Radar deception
  jamming recognition based on weighted ensemble cnn with transfer learning,''
  \emph{IEEE Trans. Geosci. Remote Sens.}, vol.~60, pp. 1--11, 2022.

\bibitem{10587212}
Y.~Wang, X.~Chen, X.~Wang, C.~Zang, W.~Zhao, and G.~Cui, ``Radar detection
  self-evolution based on incremental learning in time-varying clutter
  environments,'' \emph{IEEE Trans. Instrum. Meas.}, vol.~73, pp. 1--13, 2024.

\bibitem{10975280}
X.~Wang, Y.~Wang, S.~Guo, W.~Zhao, C.~Zang, X.~Chen, and G.~Cui, ``Intelligent
  maritime radar target detection in unknown environment via self-learning,''
  \emph{IEEE Trans. Aerosp. Electron. Syst.}, vol.~61, no.~5, pp.
  11\,172--11\,186, 2025.

\bibitem{11359362}
X.~Du, D.~Ma, X.~Chen, G.~Cui, and J.~Zheng, ``Td-amrknet-based radar image
  processing framework for sea clutter suppression,'' \emph{IEEE Trans. Radar
  Syst.}, vol.~4, pp. 387--406, 2026.

\bibitem{11124398}
C.~Zang, G.~Cui, Y.~Wang, X.~Chen, X.~Wang, Y.~Wang, and X.~Yang, ``Radar
  self-evolution detection: Two-stage knowledge transfer via
  distillation–fusion synergy,'' \emph{IEEE Trans. Aerosp. Electron. Syst.},
  vol.~61, no.~6, pp. 16\,817--16\,836, 2025.

\bibitem{10848058}
H.~Ren, R.~Zhou, L.~Zou, and H.~Tang, ``Hierarchical distribution-based
  exemplar replay for incremental sar automatic target recognition,''
  \emph{IEEE Trans. Aerosp. Electron. Syst.}, vol.~61, no.~3, pp. 6576--6588,
  2025.

\bibitem{11231088}
S.~Dang, Y.~Sun, X.~Wu, Z.~Xia, X.~Jiang, S.~Gui, and X.~Feng,
  ``Self-supervised continual learning for sar-atr: A local feature adaptation
  framework,'' \emph{IEEE Trans. Aerosp. Electron. Syst.}, vol.~62, pp.
  1378--1393, 2026.

\bibitem{8496795}
J.~Lu, A.~Liu, F.~Dong, F.~Gu, J.~Gama, and G.~Zhang, ``Learning under concept
  drift: A review,'' \emph{IEEE Trans. Knowl. Data Eng.}, vol.~31, no.~12, pp.
  2346--2363, 2019.

\bibitem{11103500}
S.~Greco, B.~Vacchetti, D.~Apiletti, and T.~Cerquitelli, ``Unsupervised concept
  drift detection from deep learning representations in real-time,'' \emph{IEEE
  Trans. Knowl. Data Eng.}, vol.~37, no.~10, pp. 6232--6245, 2025.

\bibitem{7450691}
S.~Subedi, Y.~D. Zhang, M.~G. Amin, and B.~Himed, ``Group sparsity based
  multi-target tracking in passive multi-static radar systems using
  doppler-only measurements,'' \emph{IEEE Trans. Signal Process.}, vol.~64,
  no.~14, pp. 3619--3634, 2016.

\bibitem{10066514}
D.~Song, Q.~Feng, S.~Chen, F.~Xi, and Z.~Liu, ``Space–time adaptive
  processing using deep neural network-based shrinkage algorithm under small
  training samples,'' \emph{IEEE Trans. Aerosp. Electron. Syst.}, vol.~59,
  no.~6, pp. 9697--9703, 2023.

\bibitem{9655472}
W.~Deng and L.~Zheng, ``Autoeval: Are labels always necessary for classifier
  accuracy evaluation?'' \emph{IEEE Trans. Pattern Anal. Mach. Intell.},
  vol.~46, no.~3, pp. 1868--1880, 2024.

\bibitem{yu2022predicting}
Y.~Yu, Z.~Yang, A.~Wei, Y.~Ma, and J.~Steinhardt, ``Predicting
  out-of-distribution error with the projection norm,'' in \emph{Int. Conf.
  Mach. Learn.}\hskip 1em plus 0.5em minus 0.4em\relax PMLR, 2022, pp.
  25\,721--25\,746.

\bibitem{10858421}
G.~Xu, L.~Yi, P.~Xu, J.~Li, R.~Pu, C.~Shui, A.~I. McLeod, B.~Wang, and C.~Ling,
  ``Unraveling the mysteries of label noise in source-free domain adaptation:
  Theory and practice,'' \emph{IEEE Trans. Pattern Anal. Mach. Intell.},
  vol.~47, no.~5, pp. 4007--4023, 2025.

\bibitem{jiang2022assessing}
Y.~Jiang, V.~Nagarajan, C.~Baek, and J.~Z. Kolter, ``Assessing generalization
  of {SGD} via disagreement,'' in \emph{Int. Conf. Learn. Represent.}, 2022,
  pp. 1--13.

\bibitem{pagliardini2023agree}
M.~Pagliardini, M.~Jaggi, F.~Fleuret, and S.~P. Karimireddy, ``Agree to
  disagree: Diversity through disagreement for better transferability,'' in
  \emph{Eleventh Int. Conf. Learn. Represent.}, 2023, pp. 1--13.

\bibitem{kirsch2022note}
A.~Kirsch and Y.~Gal, ``A note on ``assessing generalization of {SGD} via
  disagreement'','' \emph{Trans. Mach. Learn. Res.}, pp. 1--14, 2022.

\bibitem{ovadia2019can}
Y.~Ovadia, E.~Fertig, J.~Ren, Z.~Nado, D.~Sculley, S.~Nowozin, J.~Dillon,
  B.~Lakshminarayanan, and J.~Snoek, ``Can you trust your model's uncertainty?
  evaluating predictive uncertainty under dataset shift,'' \emph{Adv. Neural
  Inf. Process. Syst.}, vol.~32, pp. 1--12, 2019.

\bibitem{lipton2018detecting}
Z.~Lipton, Y.-X. Wang, and A.~Smola, ``Detecting and correcting for label shift
  with black box predictors,'' in \emph{Int. Conf. Mach. Learn.}\hskip 1em plus
  0.5em minus 0.4em\relax PMLR, 2018, pp. 3122--3130.

\bibitem{chen2021mandoline}
M.~Chen, K.~Goel, N.~S. Sohoni, F.~Poms, K.~Fatahalian, and C.~R{\'e},
  ``Mandoline: Model evaluation under distribution shift,'' in \emph{Int. Conf.
  Mach. Learn.}\hskip 1em plus 0.5em minus 0.4em\relax PMLR, 2021, pp.
  1617--1629.

\bibitem{donmez2010unsupervised}
P.~Donmez, G.~Lebanon, and K.~Balasubramanian, ``Unsupervised supervised
  learning i: Estimating classification and regression errors without labels.''
  \emph{J. Mach. Learn. Res.}, vol.~11, no.~4, 2010.

\bibitem{10356834}
F.~Zhu, X.-Y. Zhang, Z.~Cheng, and C.-L. Liu, ``Revisiting confidence
  estimation: Towards reliable failure prediction,'' \emph{IEEE Trans. Pattern
  Anal. Mach. Intell.}, vol.~46, no.~5, pp. 3370--3387, 2024.

\bibitem{garg2022leveraging}
S.~Garg, S.~Balakrishnan, Z.~C. Lipton, B.~Neyshabur, and H.~Sedghi,
  ``Leveraging unlabeled data to predict out-of-distribution performance,'' pp.
  1--13, 2022.

\bibitem{kiyasseh2024framework}
D.~Kiyasseh, A.~Cohen, C.~Jiang, and N.~Altieri, ``A framework for evaluating
  clinical artificial intelligence systems without ground-truth annotations,''
  \emph{Nat. Commun.}, vol.~15, no.~1, pp. 1808--1825, 2024.

\bibitem{lu2023characterizing}
Y.~Lu, Y.~Qin, R.~Zhai, A.~Shen, K.~Chen, Z.~Wang, S.~Kolouri, S.~Stepputtis,
  J.~Campbell, and K.~Sycara, ``Characterizing out-of-distribution error via
  optimal transport,'' \emph{Adv. Neural Inf. Process. Syst.}, vol.~36, pp.
  17\,602--17\,622, 2023.

\bibitem{10452835}
J.~Li, Z.~Yu, Z.~Du, L.~Zhu, and H.~T. Shen, ``A comprehensive survey on
  source-free domain adaptation,'' \emph{IEEE Trans. Pattern Anal. Mach.
  Intell.}, vol.~46, no.~8, pp. 5743--5762, 2024.

\bibitem{10476763}
A.~Khamis, R.~Tsuchida, M.~Tarek, V.~Rolland, and L.~Petersson, ``Scalable
  optimal transport methods in machine learning: A contemporary survey,''
  \emph{IEEE Trans. Pattern Anal. Mach. Intell.}, pp. 1--20, 2024.

\bibitem{gibbs2002choosing}
A.~L. Gibbs and F.~E. Su, ``On choosing and bounding probability metrics,''
  \emph{Int. Stat. Rev.}, vol.~70, no.~3, pp. 419--435, 2002.

\bibitem{peyre2019computational}
G.~Peyré and M.~Cuturi, ``Computational optimal transport: With applications
  to data science,'' \emph{Found. Trends Mach. Learn.}, vol.~11, no.~5, pp.
  355--607, 2019.

\bibitem{balasubramanian2011unsupervised}
K.~Balasubramanian, P.~Donmez, and G.~Lebanon, ``Unsupervised supervised
  learning ii: Margin-based classification without labels,'' in \emph{Proc.
  Fourteenth Int. Conf. Artif. Intell. Stat.}\hskip 1em plus 0.5em minus
  0.4em\relax JMLR Workshop and Conference Proceedings, 2011, pp. 137--145.

\bibitem{nguyen2026reliably}
V.~Nguyen, C.~Shui, V.~Giri, S.~Arya, A.~Verma, F.~Razak, and R.~Krishnan,
  ``Reliably detecting model failures in deployment without labels,''
  \emph{Adv. Neural Inf. Process. Syst.}, vol.~38, pp. 7864--7908, 2026.

\bibitem{9238468}
S.~Zhao, X.~Yue, S.~Zhang, B.~Li, H.~Zhao, B.~Wu, R.~Krishna, J.~E. Gonzalez,
  A.~L. Sangiovanni-Vincentelli, S.~A. Seshia, and K.~Keutzer, ``A review of
  single-source deep unsupervised visual domain adaptation,'' \emph{IEEE Trans.
  Neural Netw. Learn. Syst.}, vol.~33, no.~2, pp. 473--493, 2022.

\bibitem{chen2022estimating}
L.~Chen, M.~Zaharia, and J.~Y. Zou, ``Estimating and explaining model
  performance when both covariates and labels shift,'' \emph{Adv. Neural Inf.
  Process. Syst.}, vol.~35, pp. 11\,467--11\,479, 2022.

\bibitem{rousseeuw2011robust}
P.~J. Rousseeuw and M.~Hubert, ``Robust statistics for outlier detection,''
  \emph{Wiley Interdiscip. Rev. Data Min. Knowl. Discov.}, vol.~1, no.~1, pp.
  73--79, 2011.

\bibitem{jian2023sea}
G.~Jian, L.~Ningbo, W.~Guoqing, D.~Hao, D.~Yunlong, H.~Yong, T.~Kaixiang, and
  Z.~Mengyu, ``Sea-detecting radar experiment and target feature data
  acquisition for dual polarization multistate scattering dataset of marine
  targets,'' \emph{J. Radars}, vol.~12, no.~2, pp. 456--469, 2023.

\bibitem{9143512}
Z.~Chen, L.~Deng, B.~Wang, G.~Li, and Y.~Xie, ``A comprehensive and modularized
  statistical framework for gradient norm equality in deep neural networks,''
  \emph{IEEE Trans. Pattern Anal. Mach. Intell.}, vol.~44, no.~1, pp. 13--31,
  2022.

\bibitem{guillory2021predicting}
D.~Guillory, V.~Shankar, S.~Ebrahimi, T.~Darrell, and L.~Schmidt, ``Predicting
  with confidence on unseen distributions,'' in \emph{Proc. IEEE/CVF Int. Conf.
  Comput. Vis.}, 2021, pp. 1134--1144.

\bibitem{liu2015classification}
T.~Liu and D.~Tao, ``Classification with noisy labels by importance
  reweighting,'' \emph{IEEE Trans. Pattern Anal. Mach. Intell.}, vol.~38,
  no.~3, pp. 447--461, 2015.

\bibitem{villani2009optimal}
C.~Villani, \emph{Optimal transport: old and new}, ser. Grundlehren der
  mathematischen Wissenschaften.\hskip 1em plus 0.5em minus 0.4em\relax Berlin,
  Heidelberg: Springer, 2009, vol. 338, pp. 72--95.

\end{thebibliography}
\end{document}